\documentclass[11pt, a4paper]{article}
\pdfoutput=1

\usepackage{anysize}
\marginsize{3cm}{3cm}{1.5cm}{1.5cm}
\usepackage[T1]{fontenc}
\usepackage{inputenc}

\usepackage{amsfonts}
\usepackage{amssymb}
\usepackage{graphics}
\usepackage{epsfig}
\usepackage{graphicx}
\usepackage{epsfig}
\usepackage{amssymb}
\usepackage{amsfonts}

\usepackage{amsmath}
\usepackage{xcolor}
\usepackage{float}
\usepackage{enumerate}
\usepackage{slashed}
\usepackage{hyperref}
\hypersetup{colorlinks=true,linkcolor=blue,citecolor=blue}

\usepackage{caption}
\usepackage{subcaption}

\numberwithin{equation}{section}

\newcommand {\beq} {\begin{equation}}
\newcommand {\eeq} {\end{equation}}
\newcommand{\bea}{\begin{eqnarray}}
\newcommand{\eea}{\end{eqnarray}}
\newcommand{\bit}{\begin{itemize}}
\newcommand{\eit}{\end{itemize}}
\def\nl{\nonumber \\}

\def\l{\lambda}

\def\s{\sigma}

\def\p{\partial}

\def\le{\left(}
\def\ri{\right)}

\def\beq{\begin{equation}}
\def\eeq{\end{equation}}

\newcommand{\Eq}[1]{(\ref{#1})}

\begin{document}

\begin{titlepage}

\begin{flushright}

\end{flushright}
\bigskip
\begin{center}
{\LARGE  {\bf
On volume subregion complexity 
in Vaidya  spacetime
  \\[2mm] } }
\end{center}
\bigskip
\begin{center}
{\large \bf  Roberto  Auzzi$^{a,b}$},
  {\large \bf Giuseppe Nardelli$^{a,c}$},  \\
  {\large \bf Fidel I. Schaposnik Massolo$^{d}$},
    {\large \bf Gianni Tallarita$^{e}$}, 
    {\large \bf Nicol\`o Zenoni$^{a,b,f}$}
\vskip 0.20cm
\end{center}
\vskip 0.20cm 
\begin{center}
$^a${ \it \small  Dipartimento di Matematica e Fisica,  Universit\`a Cattolica
del Sacro Cuore, \\
Via Musei 41, 25121 Brescia, Italy}
\\ \vskip 0.20cm 
$^b${ \it \small{INFN Sezione di Perugia,  Via A. Pascoli, 06123 Perugia, Italy}}
\\ \vskip 0.20cm 
$^c${ \it \small{TIFPA - INFN, c/o Dipartimento di Fisica, Universit\`a di Trento, \\ 38123 Povo (TN), Italy} }
\\ \vskip 0.20cm 
$^d${ \it Institut des Hautes \'Etudes Scientifiques,\\35 route de Chartres, 91440 Bures-sur-Yvette, France}
\\ \vskip 0.20cm 
$^e${ \it \small{Departamento de Ciencias, Facultad de Artes Liberales,
 Universidad Adolfo Ib\'a\~nez, 
Santiago 7941169, Chile,}}
\\ \vskip 0.20cm 
$^f${ \it \small{ 
Instituut voor Theoretische Fysica, KU Leuven, Celestijnenlaan 200D, B-3001 Leuven, Belgium } }
\\ \vskip 0.20cm 
E-mails: {\tt roberto.auzzi@unicatt.it, giuseppe.nardelli@unicatt.it, fidels@ihes.fr, gianni.tallarita@uai.cl,
nicolo.zenoni@unicatt.it}
\end{center}
\vspace{3mm}

\begin{abstract}
We study holographic subregion volume complexity
for a line segment in the AdS$_3$ Vaidya geometry.
 On the field theory side, this gravity background corresponds to a sudden quench which leads 
 to the thermalization of the strongly-coupled dual conformal field theory.
 We find the time-dependent extremal volume surface
by numerically solving a partial differential equation
with boundary condition given by the Hubeny-Rangamani-Takayanagi  
surface, and we use this solution to compute holographic subregion complexity
as a function of time. Approximate analytical expressions
 valid at early and at late times are derived.
\end{abstract}

\end{titlepage}

\section{Introduction}

In the AdS/CFT correspondence,
quantum information concepts such as entanglement entropy have 
simple geometrical descriptions, e.g. the area of a minimal surface
in the bulk gravity dual \cite{Ryu:2006bv,Casini:2011kv,Lewkowycz:2013nqa}.
These results put on a more general picture the idea that
the area of the event
 horizon is proportional to the black hole entropy \cite{Bekenstein:1973ur}.
It is then reasonable that more sophisticated quantum information 
physical quantities computed on the boundary theory may give 
us further insights on how other geometrical properties 
 of the bulk dual may be reconstructed from the boundary.

Recently, a new quantum information concept has been introduced in order to describe
the growth of the Einstein-Rosen Bridge (ERB) inside the horizon of a 
black hole, which continues for a much longer time than the thermalization time.
Entanglement entropy is not enough to describe the 
dynamics behind the event horizon and the late-time evolution of the wormhole interior,
because it approaches the equilibrium on a time scale
 which is of  the same order as the thermalization  time scale.
It has been suggested that the relevant quantity in the dual field theory is 
quantum computational complexity  \cite{Susskind:2014rva,Stanford:2014jda,Susskind:2014moa}.
This is heuristically defined as the minimal number of elementary unitary operations that are required
in order to prepare a given state from a reference one.
In quantum mechanics, a geometrical approach to complexity
 was developed by Nielsen and collaborators \cite{Nielsen1,Nielsen2}.
 In Quantum Field Theory (QFT), a rigorous definition of complexity 
 involves several subtleties, see e.g.
  \cite{Jefferson:2017sdb,Chapman:2017rqy,Hashimoto:2017fga,Caputa:2017yrh,Bhattacharyya:2018wym,Chapman:2018hou}
for attempts to define it more rigorously.

Two holographic quantities have been conjectured to be the gravity dual of complexity: 
\begin{itemize}
\item Complexity=Volume (CV) conjecture: complexity is proportional to
the volume of extremal space-like slices \cite{Susskind:2014rva,Stanford:2014jda,Susskind:2014moa};
\item Complexity=Action (CA) conjecture: complexity is proportional to the action 
evaluated on the Wheeler-deWitt (WdW) patch \cite{Brown:2015bva,Brown:2015lvg}.
It is interesting that a proper action calculation involves 
null boundaries and joint  terms
that have been recently studied in \cite{Lehner:2016vdi}.
\end{itemize}
Both conjectures have been recently investigated by several groups in many physical settings, e.g. 
\cite{Cai:2016xho,Chapman:2016hwi,Carmi:2017jqz,Alishahiha:2017hwg,Ghodrati:2017roz,
Auzzi:2018zdu,Auzzi:2018pbc,Dimov:2019fxp,Alishahiha:2018tep,Flory:2018akz,Flory:2019kah}.
One interesting situation is the global quench, which can be represented
in AdS/CFT by the Vaidya geometry, see e.g. \cite{Balasubramanian:2011ur}.
Holographic complexity in these geometries was previously studied in 
\cite{Moosa:2017yvt,Moosa:2017yiz,Chapman:2018dem,Chapman:2018lsv}.
Another interesting situation is the local quench \cite{Nozaki:2013wia},
whose complexity was studied in \cite{Ageev:2018nye,Ageev:2019fxn}.

Quantum states localised on a subregion on the boundary
should be dual to the entanglement wedge \cite{Czech:2012bh,Hubeny:2012wa}. 
Consequently,  it is natural to conjecture
that the complexity of a  mixed state (which should be properly defined)
is dual to some version of the holographic CV or CA conjecture, adapted 
to the corresponding  subregion \cite{Alishahiha:2015rta,Carmi:2016wjl}.

For the CV proposal, it is natural to conjecture
\cite{Alishahiha:2015rta} that such mixed state complexity
is dual to the extremal volume of the region $\gamma$ delimited by the boundary subregion on which the
mixed state is localised and its Hubeny-Rangamani-Takayanagi (HRT) \cite{Hubeny:2007xt} surface,
whose area corresponds to the holographic entanglement entropy, i.e.
\beq
C_V=\max_\gamma\frac{V(\gamma)}{ G L_{AdS}} \, ,
\eeq
where $G$ is the Newton constant and $L_{AdS}$ the AdS length scale.
Concerning the CA conjecture, a proposal involving the action defined on a region $\Sigma$
which is the intersection of the entanglement wedge and of the WdW patch
has been introduced in \cite{Carmi:2016wjl},
\beq
C_A= \frac{I_\Sigma}{\pi \hbar}\ .
\eeq
In both cases, the precise nature of the conjectured notion of
mixed state complexity is still unknown 
and several proposals have been put forward, see e.g.
\cite{Alishahiha:2015rta,Agon:2018zso,Caceres:2018blh}.
Other studies on subregion complexity include
\cite{Ben-Ami:2016qex,Abt:2017pmf,Abt:2018ywl,Alishahiha:2018lfv,
Roy:2017kha,Roy:2017uar,Bakhshaei:2017qud,Bhattacharya:2019zkb,Auzzi:2019fnp}.

In order to get insights on the possible field theory dual quantities,
 it is necessary to explicitly compute subregion complexity in several physical settings.
The purpose of this paper is to study holographic subregion volume complexity,
using the CV conjecture, for a line segment in
 the AdS$_3$ Vaidya spacetime.
The study of subregion complexity in this physical situation
 was initiated in \cite{Chen:2018mcc}.
Moreover, the issue was studied also in modified gravity
\cite{Ling:2018xpc,Zhou:2019jlh}.
In all these previous works, an ansatz in which the
extremal volume is taken independent of the spatial coordinate
$x$ is used. This is correct in the case of time-independent geometries;
however we find that this ansatz is not consistent with the 
boundary condition given by the HRT surface for the Vaidya geometry.
In this paper we determine the extremal surface numerically and we find
that the $x$-independent ansatz is in general a good approximation 
only at early and late times. 
In the case of small subregion size $l \ll 1/T$, where $T$ is the temperature, 
the $x$-independent ansatz provides a good approximation also at intermediate times.

The paper is organised as follows: in section \ref{sect:geo}
we review the analytic solution for the HRT surfaces in case
of zero thickness shell. In section \ref{sect:volume}
we show that the $x$-independent ansatz is not consistent 
for the extremal volume in the
time-dependent case  and we compute
the $x$-dependent solution and its volume numerically.
We conclude in section \ref{sect:conclu}.
Some technical details are collected in appendices.

 \bigskip
{\bf Note added:} After this work was finished and the present
paper was in the writing stage, 
Ref.   \cite{Ling:2019ien} was submitted on arXiv.
Our approximate analytical results agree with them in the
early  time regime. At intermediate times,
we expect that  the $x$-independent 
ansatz used in    \cite{Ling:2019ien} is not accurate.

\section{Space-like geodesics}
\label{sect:geo}

We study the Complexity=Volume conjecture for subregions in AdS$_3$ Vaidya spacetime. 
In three dimensions, the HRT surface attached to a segment coincides with a space-like
geodesic. Here we review some basic aspects of these geodesics following  \cite{Balasubramanian:2011ur},
which studies the thermalization of the  entanglement entropy in detail.
We use interchangeably $r$ or $z=1/r$ as a radial AdS coordinate.
The spacetime metric is 
\bea
\label{met1}
ds^{2} &=& -r^{2} f(v,r) \, dv^{2} +2 \, dv \, dr + r^{2} \, dx^{2}  \nl
&=& \frac{1}{z^{2}}\left[ -f\left( v,z \right) dv^{2}-2 \, dv \, dz+dx^{2} \right]\,,
\eea
where we have fixed the AdS radius $L_{AdS}=1$ and
\begin{equation}
f=1-\frac{m(v)}{r^2} =1-m(v) \, z^{2}   \, .
\end{equation}

The $v$ coordinate is constant along infalling null rays  and it coincides with the time coordinate $t$ 
on the spacetime boundary,
 located at $r \rightarrow \infty$ (or, equivalently, at $z \rightarrow 0$). 
 For constant $m(v)$, changing variables to $t$, with $dv=dt-\frac{dz}{f}$, the 
solution is the Banados-Teitelboim-Zanelli (BTZ) \cite{Banados:1992wn,Banados:1992gq}
 black hole in Schwarzschild coordinates:
\beq
ds^2=\frac{1}{z^{2}} \le -f dt^2 +\frac{dz^2}{f} + dx^2\ri \, .
\eeq
We will be interested in the case in which 
the function $m(v)$ models a field theory quench, i.e. it
interpolates between $m=0$  and $m=M$. 

For concreteness, in the numerical calculations we will consider the choice
\beq
\label{mass-shell}
\qquad m(v)=\frac{M}{2} \le 1+\tanh \frac{v}{\tilde{v}} \ri \, ,
\eeq
where $M$ is proportional to the final BH mass and $\tilde{v} $ parameterizes the thickness of the shell.
The $\tilde{v} \rightarrow 0$ limit corresponds to zero thickness;
in this case $m(v)$ can be written in terms of the Heaviside step function  $\vartheta$:
\begin{equation}
\label{heaviside}
m(v)= M \, \vartheta (v) \, .
\end{equation}
In the zero thickness limit, analytical expressions for the geodesics are available.
With the choice (\ref{heaviside}), the geometry described by eq. \Eq{met1} is the  AdS$_{3}$ one for $v<0$ 
and the BTZ
 black hole \cite{Banados:1992wn,Banados:1992gq} one for $v>0$. 
The BTZ black hole is
 formed by the gravitational collapse of a shell of null dust
  (here described by $v=0$) with infinitesimal thickness  falling from the spacetime boundary.

Our purpose is to evaluate the subregion complexity of a boundary subregion. 
According to the CV conjecture for subregions, we have to compute 
the volume of an extremal codimension-one bulk surface delimited by the boundary
 subregion and the corresponding codimension-two HRT  \cite{Hubeny:2007xt}  surface. 
 In the $2+1$ dimensional case, the $1$-dimensional HRT surface 
 is a space-like geodesic anchored at the edges of the boundary subregion. 

We consider as a subregion a segment of length $l$ lying on a constant time slice $t$ 
on the boundary, described by $x \in \left[ -\frac{l}{2}, \frac{l}{2} \right]$. 
The HRT surface can be parameterized as $v(x)$ and $r(x)$. 
The boundary conditions at $r=\infty$ are
\beq
\label{bc1}
x(r=\infty)= \pm \frac{l}{2} \, , \qquad v(r=\infty)= t \,.
\eeq
By symmetry, the turning point is at $x=0$, i.e.
\beq
x(r=r_*)=0 \, , \qquad v(r=r_*)=v_* \, ,
\eeq
where $r_*$ denotes the value of $r$ at the turning point. 
Note that both $r_*$ and $v_*$ are functions of the geodesic boundary condition $t$.

 Since the  spacetime is  
described by  an  AdS$_{3}$ part and a BTZ black hole portion glued at $v=0$,
the HRT surface is given by the junction at $v=0$ of the HRT surface for a BTZ
spacetime and the one for AdS$_{3}$ spacetime.\footnote{We consider the general case
 in which the HRT surface crosses the infalling shell of matter.} 
 In the following we denote with $r=r_{s}$ the position of this junction on the $v=0$ infalling null ray. 

\subsection{AdS$_3$ geodesics}

For $v<0$, the Vaidya spacetime  is   AdS$_{3}$:
\begin{equation}
\label{btz-metric}
ds^{2} = -r^{2}  dv^{2} +2 \, dv \, dr + r^{2} \, dx^{2}\ .
\end{equation}
The corresponding portion of the HRT surface is given by the equal-time
space-like geodesic in the  AdS geometry:
\begin{equation}
\label{geo-ads}
x_\pm(r)=\pm \frac{\sqrt{r^2-r_*^2}}{r_* r} \, ,
\qquad 
v_\pm(r)= \frac{1}{r_s}-\frac{1}{r} \, ,
\eeq
where $(r_*,r_s)$ are functions of the boundary time $t$ and of the length $l$.
We will denote  $\left( x_{+}(r), v_{+}(r) \right)$ and $\left( x_{-}(r), v_{-}(r) \right)$ 
as branches 1 and 2 of the geodesic, respectively. 
At initial time $t=0$, the geodesic is entirely in AdS and
\beq
r_*(t=0)
=\frac{2}{l} \, .
\end{equation}

\subsection{BTZ geodesics}

For $v>0$, the Vaidya spacetime is a BTZ black hole:
\begin{equation}
\label{btz-metric}
ds^{2} = -r^{2} \left( 1- \frac{r_h^2}{r^{2}} \right) dv^{2} +2 \, dv \, dr + r^{2} \, dx^{2}\ .
\end{equation}
The event horizon of the black hole is located at $r=r_{h}$ and
the Hawking temperature is $T=\frac{r_h}{2 \pi}$.

\begin{figure}[h]
\center
\includegraphics[scale=0.8]{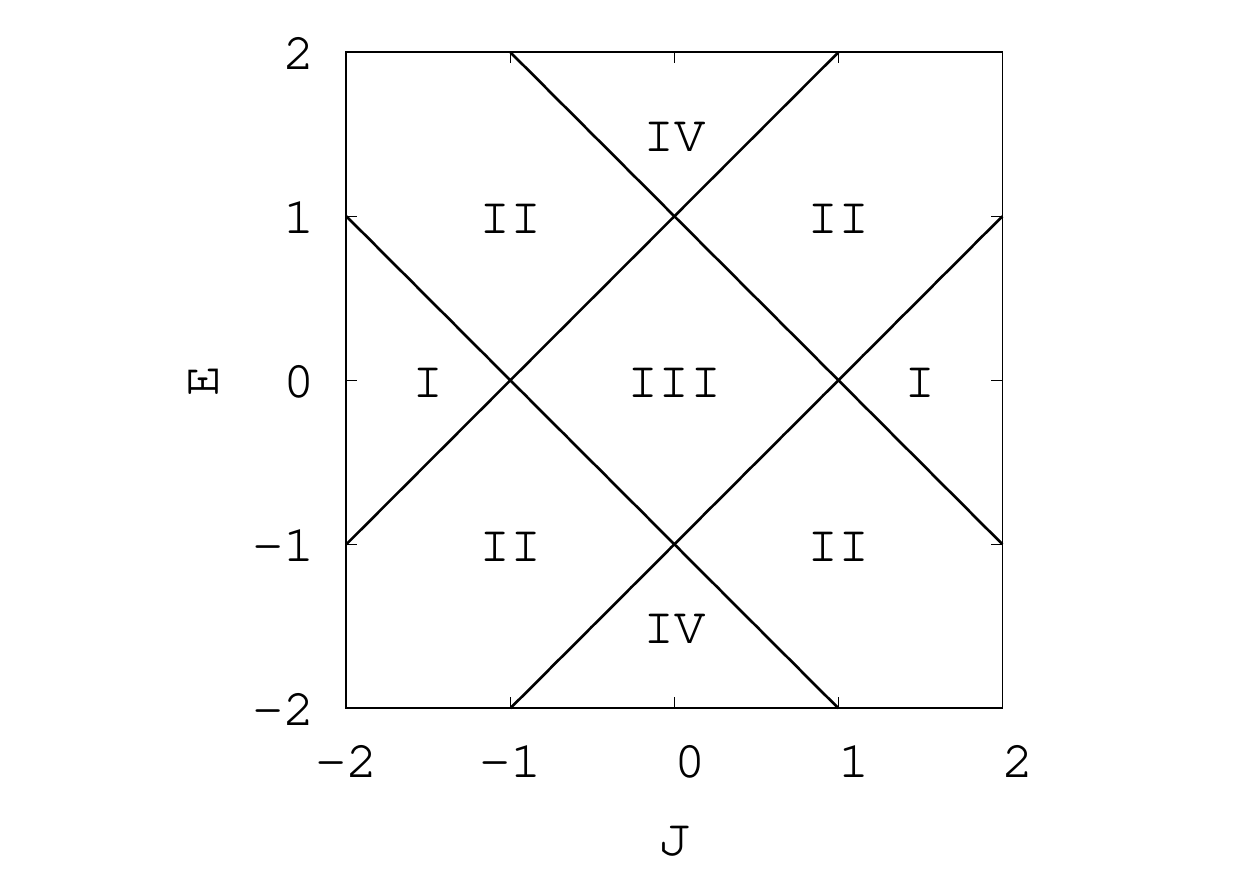}
\caption{Kinds of space-like geodesics as function of $(J,E)$. }
\label{regions}
\end{figure}

The part of the HRT surface in the Vaidya spacetime for $v>0$ is given by the 
space-like geodesic in the BTZ geometry \cite{Balasubramanian:2011ur}:
\begin{equation}
\label{geo1}
x_{\pm}(r)= \frac{1}{4 r_{h}} \left\{ 2 \ln 
\left| \frac{r^{2}-J \, r_{h}^{2} \pm \sqrt{r^{4}+\left( E^{2}-J^{2} -1\right) r_{h}^{2} \, r^{2}+ J^{2} \, r_{h}^{4}}}{r^{2}+J \, r_{h}^{2} \pm \sqrt{r^{4}+\left( E^{2}-J^{2}-1 \right) r_{h}^{2} \, r^{2}+ J^{2} \, r_{h}^{4}}} \right|
+\ln \left| \frac{(J+1)^{2} -E^{2}}{(J-1)^{2}-E^{2}} \right|  \right\}\, ,
\end{equation}
\begin{equation}
\label{geo2}
v_{\pm}(r)= t+ \frac{1}{2 r_{h}}
 \ln \left| \frac{r-r_{h}}{r+r_{h}}\  \frac{r^{2}-\left( E+1 \right)  r_{h}^{2} \pm \sqrt{r^{4}+\left( E^{2}-J^{2} -1\right) r_{h}^{2} \, r^{2}+ J^{2} \, r_{h}^{4}}}{r^{2}+ \left( E-1 \right)  r_{h}^{2} \pm \sqrt{r^{4}+\left( E^{2}-J^{2} -1 \right) r_{h}^{2} \, r^{2}+ J^{2} \, r_{h}^{4}}} \right|  \, ,
\end{equation}
with $E$ and $J$ being two integration constants arising  from the equations of motion
 (see appendix \ref{Appe-btz-geo}). 
Depending on the values of $E,J$ in (\ref{geo1}), (\ref{geo2}),  the structure 
of the geodesic changes; it is useful to distinguish four regions  \cite{Balasubramanian:2011ur},
see Fig. \ref{regions}.
In our notation, we have translated the solutions in $x$  
in such a way that
 they are symmetric  under the exchange $x \rightarrow -x$.

\begin{figure}[h]
\center
\begin{tabular}{cc}
\includegraphics[width=7cm]{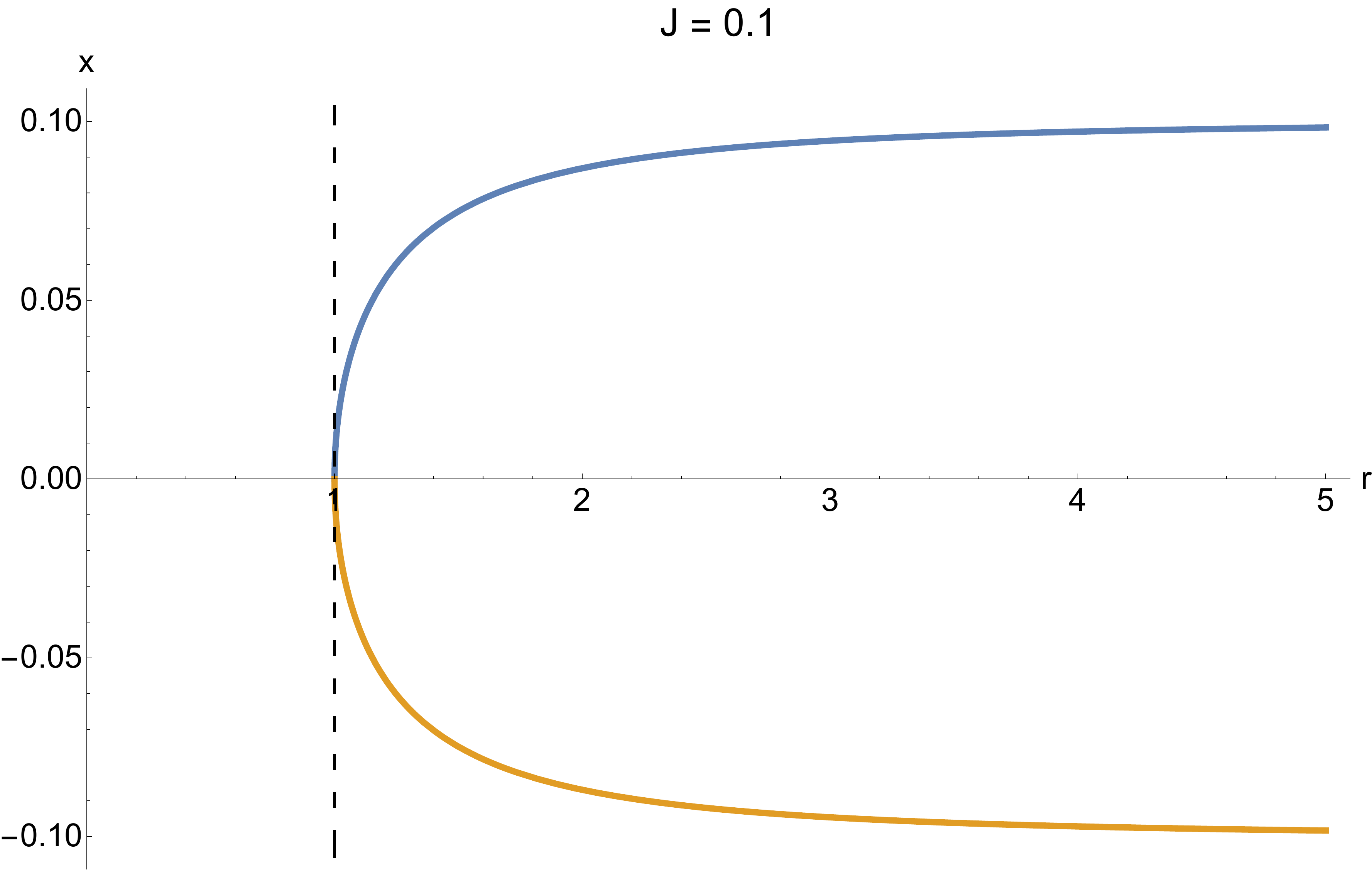} & \includegraphics[width=7cm]{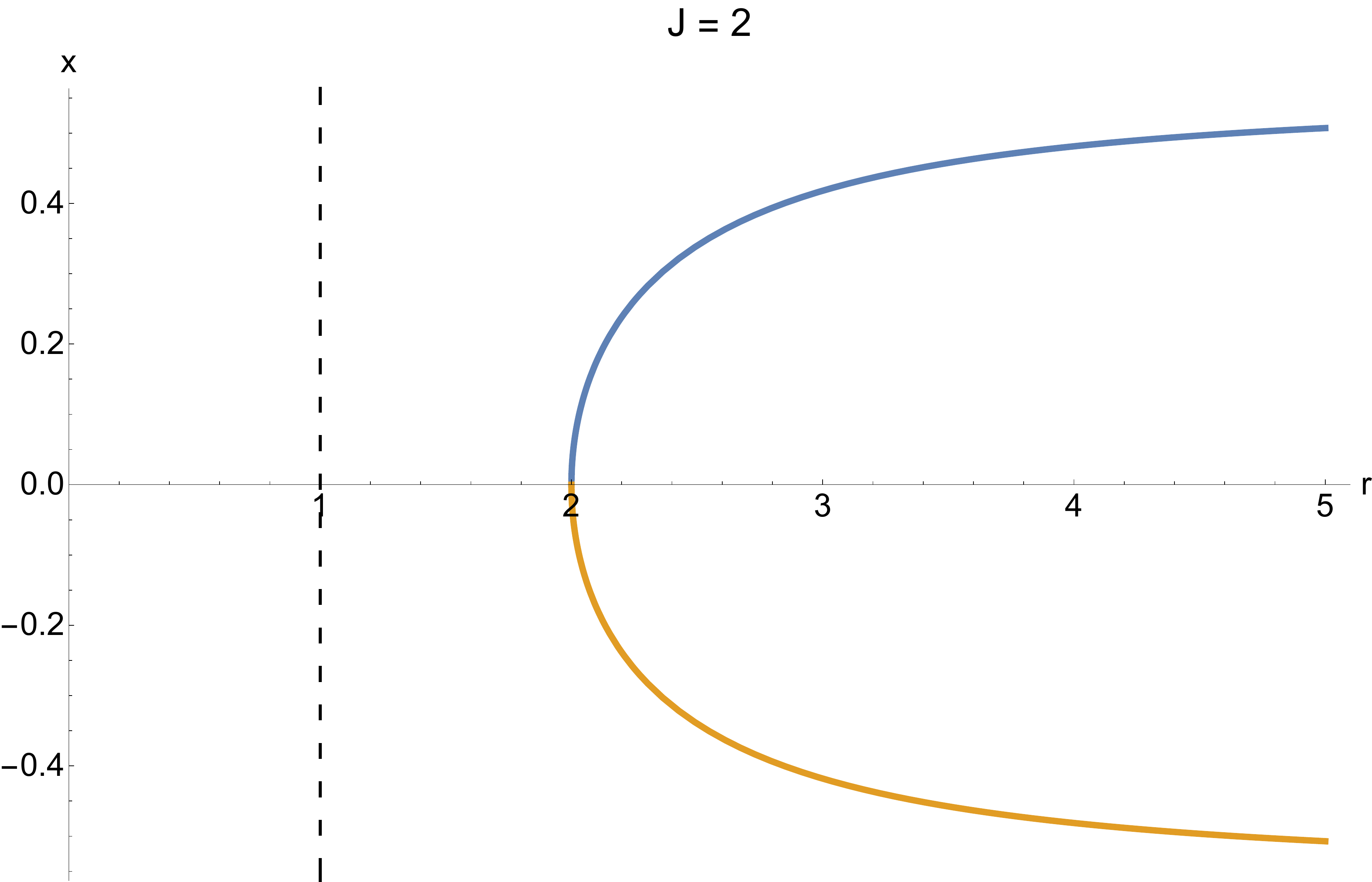}
\end{tabular}
\caption{Plots of the space-like geodesic \Eq{geo1} in BTZ spacetime with $E=0$ 
and different values of the parameter $J$, with $r_{h}=1$. 
The blue curve represents $x_{+}(r)$, while the yellow one represents  $x_{-}(r)$.}
\label{fig:fig2}
\end{figure}

Let us start for simplicity with $E=0$, which corresponds to geodesics  lying 
on $t$-constant slices.  By symmetry, it is not restrictive to choose $J>0$ and then
 there are only two kinds of such geodesics (see Fig. \ref{regions}):
 the ones with $J>1$ (region $I$) and the ones with $J<1$
 (region $III$)\footnote{In the special case $E=0$ and $J=1$,  the geodesic is singular.
  We shall see that this value will be never attained in our context.}.
In Fig. \ref{fig:fig2} we show the plot of the geodesic \Eq{geo1} for both the cases $J>1$ and $J<1$. 
By direct calculation, we find that the minimal value of $r$ along the geodesic is:
\beq
 r_0 = 
\begin{cases}
J \, r_{h} \, , \qquad J>1 \\
r_{h} \, , \qquad \,\,\,\, J<1  \, .\\
\end{cases}
\eeq
The geodesics relevant as HRT surfaces for the static BTZ black hole
are the ones in region $I$, because they have minimal length compared to the ones in region $III$.
Note that a space-like geodesic with $E=0$
  in a static BTZ spacetime never penetrates inside the black hole.
  For $J>1$,
the relation between the  parameter $J$ and the spatial separation
 $l$ between the anchoring points of the geodesic is given by:
\begin{equation}
\label{cothJ}
\frac{l}{2} = \frac{1}{4 r_{h}} \ln \left( \frac{J+1}{J-1} \right)^{2} \, , \qquad {\rm or} \qquad
J = \coth \left( \frac{r_{h} \, l}{2} \right) \, .
\end{equation}
This allows to express $r_{0}$ as a function of the boundary separation $l$ in the $J>1$ case:
\beq
\label{coth}
r_{0}= r_{h} \coth \left( \frac{r_{h} \, l}{2} \right)\ .
\eeq

\begin{figure}[h]
\center
\begin{tabular}{cc}
\includegraphics[width=7cm]{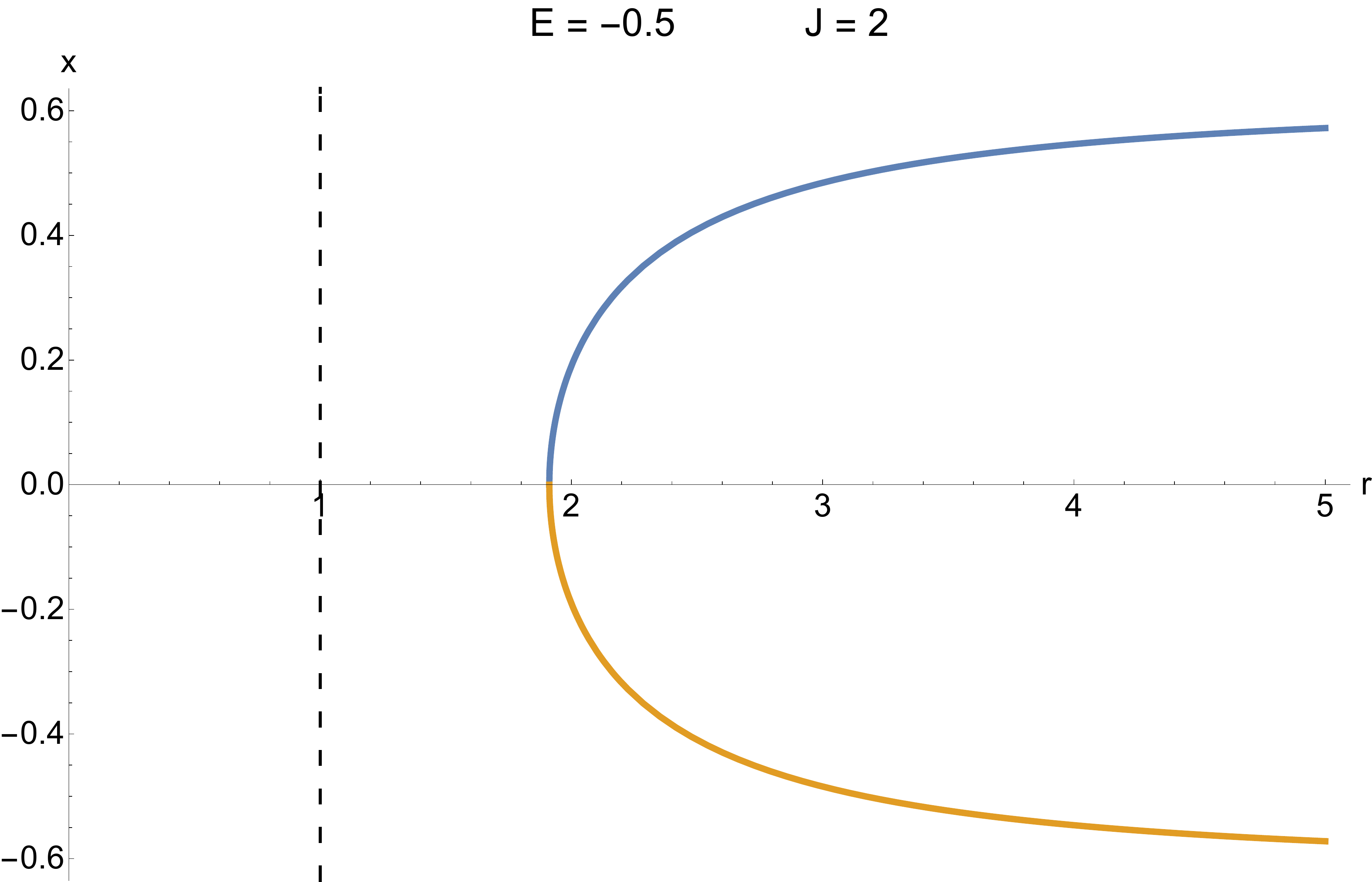} & \includegraphics[width=7cm]{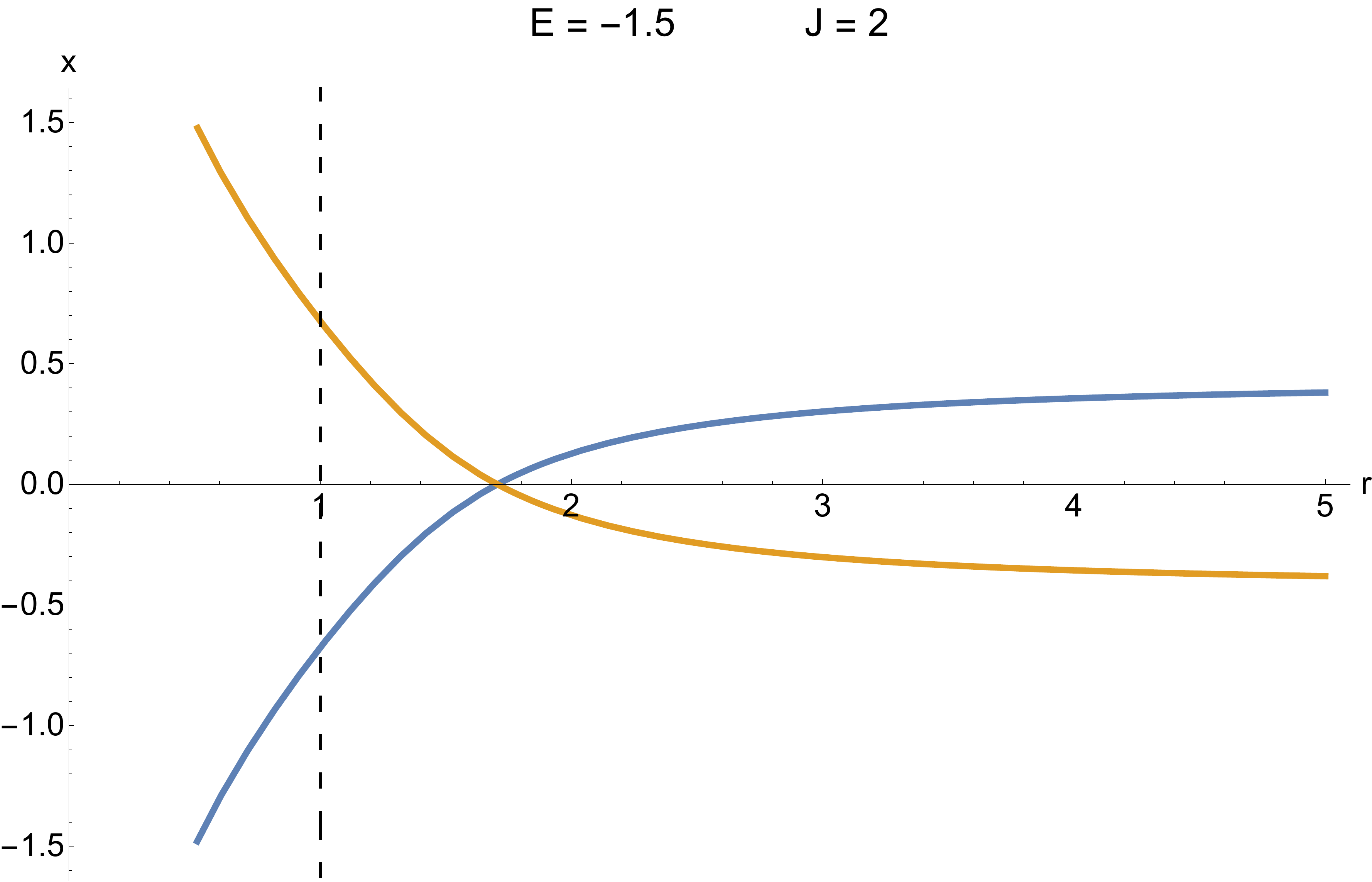}\\
(I) & (II)\\[10pt]
\includegraphics[width=7cm]{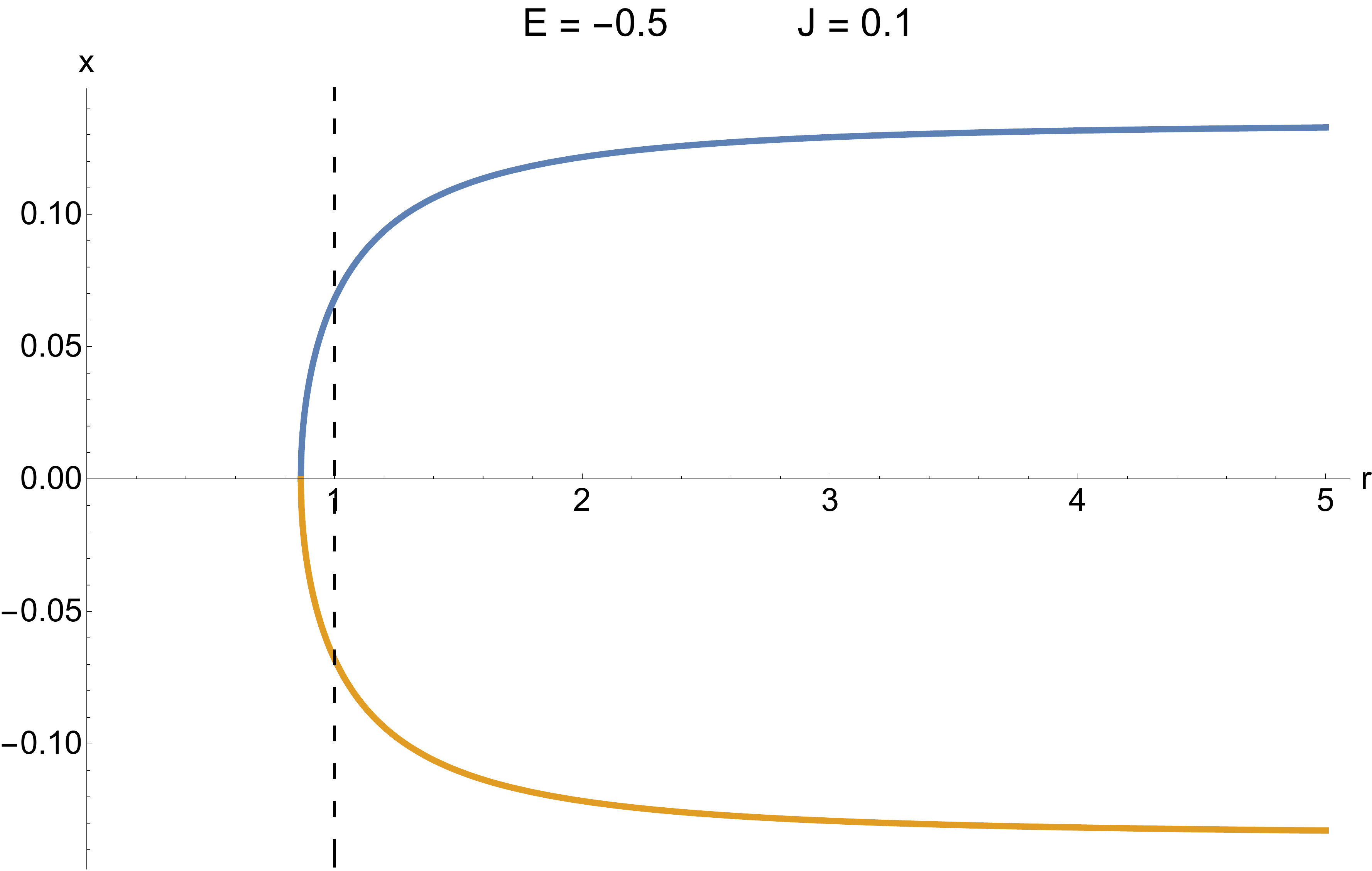} & \includegraphics[width=7cm]{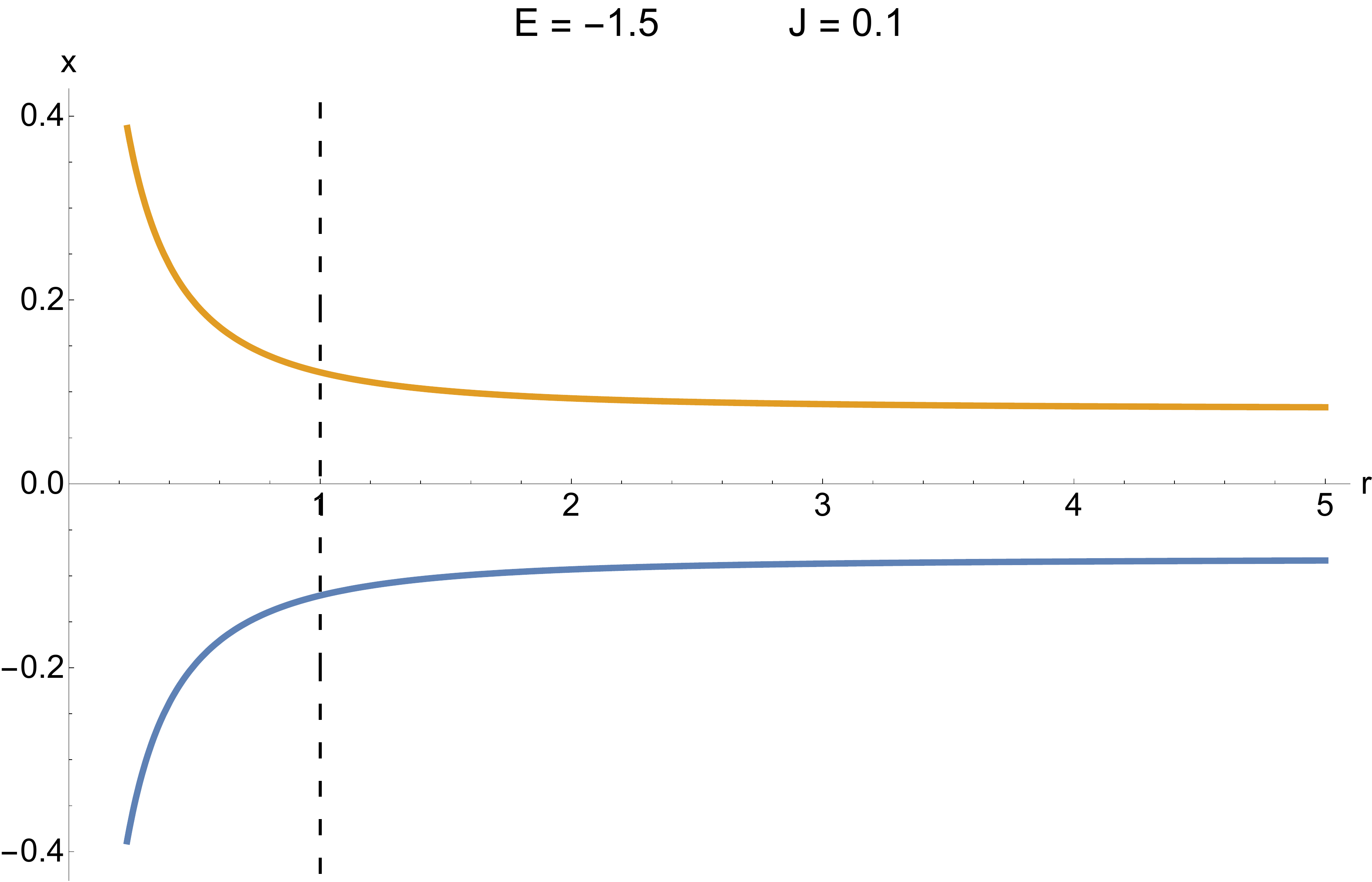}\\
(III) & (IV)
\end{tabular}
\caption{Plots of the space-like geodesic \Eq{geo1} in BTZ spacetime with different values of the parameters $(E,J)$, with $r_h=1$.
 The blue curve represents  $x_{+}(r)$, while the yellow one represents  $x_{-}(r)$.}
\label{fig:fig3}
\end{figure}

For generic $E$, there are in principle 
 four different kinds of geodesics, one kind for each region of the $(E,J)$ plane
 in Fig. \ref{regions}. In Fig. \ref{fig:fig3} we show a plot of \Eq{geo1} for each kind of geodesic.
For $E \neq 0$, these geodesics connect  points on the boundary with different values of $t$.
 Note that  the geodesic on the bottom left of Fig. \ref{fig:fig3} penetrates inside the black hole, 
 while this never happens for geodesics at constant $t$.  
 
\subsection{Joining the geodesics}
\label{sect-joining}

The HRT surface in the full Vaidya spacetime can be obtained by gluing together the AdS$_{3}$
 geodesic \Eq{geo-ads} and the BTZ one (\ref{geo1}, \ref{geo2}) at $r=r_{s}$.
 Using the "refraction-like" law  in \cite{Balasubramanian:2011ur}, 
  we can fix the two constants of motion of the BTZ portion of the geodesic:
\begin{equation}
E= - \frac{r_{h} \sqrt{r_{s}^{2}-r_{*}^{2}}}{2r_{s}^{2}} \, , \qquad J= \frac{r_{*}}{r_{h}} \, .
\label{motion-constants}
\end{equation}
 It is important to note that $E,J$ all depend on the boundary time $t$ and on the length $l$.
Let us denote by  $r_{m}$
 the minimal value of the $r$-coordinate of the BTZ portion.
If $r_{s} \geq r_{h}/\sqrt{2}$ we have to consider only branch 1, 
while if $r_{s} \leq r_{h}/\sqrt{2}$ also branch 2 comes into play. 
In the latter case, a part of branch 2 (with $r_{m} \leq r \leq r_{s}$) 
connects the AdS$_{3}$ geodesic and the full branch 1, 
 which is anchored at the spacetime boundary.
 
It is useful to define:
 \bea
 \nu^{\pm}(r)&=& t+ \frac{1}{2 r_{h}}
 \ln \left| \frac{r-r_{h}}{r+r_{h}}\  \frac{r^{2}-\left( E+1 \right)  r_{h}^{2} \pm \sqrt{r^{4}+\left( E^{2}-J^{2} -1\right) r_{h}^{2} \, r^{2}+ J^{2} \, r_{h}^{4}}}{r^{2}+ \left( E-1 \right)  r_{h}^{2} \pm \sqrt{r^{4}+\left( E^{2}-J^{2} -1 \right) r_{h}^{2} \, r^{2}+ J^{2} \, r_{h}^{4}}} \right|  \, , \nl 
 \chi^{\pm}(r)&=&
 \left( \frac{1}{2r_{h}} \ln \left| \frac{r^{2}-J \, r_{h}^{2} \pm \sqrt{r^{4}+\left( -1+E^{2}-J^{2} \right) r_{h}^{2} \, r^{2}+ J^{2} \, r_{h}^{4}}}{r^{2}+J \, r_{h}^{2} \pm \sqrt{r^{4}+\left( -1+E^{2}-J^{2} \right) r_{h}^{2} \, r^{2}+ J^{2} \, r_{h}^{4}}}\right|  + \frac{l}{2}\right) 
 \, ,
 \eea
in which the values of  $E$ and $J$ are given by eq. \Eq{motion-constants};
for fixed length $l$,  we must obey the following constraint for the quantities $(r_s,r_*)$:
\begin{equation}
\label{ll-eq}
0=2 \frac{\sqrt{r_{s}^{2}-r_{*}^{2}}}{r_{s} \, r_{*}} + \frac{1}{r_{h}} \ln \frac{2r_{s} \left( r_{s}^{2}+ r_{*} \, r_{h} \right) + \left( 2 r_{s}^{2}- r_{h}^{2} \right) \sqrt{r_{s}^{2}- r_{*}^{2}}}{2r_{s} \left( r_{s}^{2}- r_{*} \, r_{h} \right) + \left( 2 r_{s}^{2}- r_{h}^{2} \right) \sqrt{r_{s}^{2}- r_{*}^{2}}} -l=f_l(r_s,r_*)\, .
\end{equation}

 We can now build the total Vaidya geodesic by suitably glueing
 BTZ and AdS portions.
In the case $r_{s} \geq \frac{r_{h}}{\sqrt{2}}$, the total geodesic is given by:
\bea
\label{full-geo1}
\hat{x}_{\pm} (r) &=& 
\begin{cases}
\pm \frac{\sqrt{r^{2}-r_{*}^{2}}}{r_{*} \, r} 
&  {\rm if} \quad r \leq r_{s},  \,\, v \leq 0 \\
\pm   \chi^+(r)
&  {\rm if}  \quad r > r_{s}, \,\,  v>0 \, ,
\end{cases}
\nl
\hat{v}_{\pm} (r) &=& 
\begin{cases}
\frac{1}{r_s}-\frac{1}{r}
&  {\rm if} \quad r \leq r_{s},  \,\, v \leq 0 \\
  \nu^+(r)
&  {\rm if}  \quad r > r_{s}, \,\,  v>0 \, .
\end{cases}
\eea
Instead, in the case $r_{s} \leq \frac{r_{h}}{\sqrt{2}}$, the full geodesic is:
\bea
\label{full-geo2}
\hat{x}_{\pm} (r) &=& 
\begin{cases}
\pm \frac{\sqrt{r^{2}-r_{*}^{2}}}{r_{*} \, r} 
&  {\rm if} \quad r \leq r_{s}, \,\,  v \leq 0 \\
\pm \chi^+(r) 
&  {\rm if} \quad r \geq r_{m}, \,\, v>0 \\
\pm \chi^-(r)  
&  {\rm if} \quad r_{m} \leq r < r_{s}, \,\,  v>0 \, ,
\end{cases}
\nl
\hat{v}_{\pm} (r) &=& 
\begin{cases}
\frac{1}{r_s}-\frac{1}{r}
&  {\rm if} \quad r \leq r_{s}, \,\,  v \leq 0 \\
\nu^+(r) 
&  {\rm if} \quad r \geq r_{m}, \,\, v>0 \\
\nu^-(r)  
&  {\rm if} \quad r_{m} \leq r < r_{s}, \,\,  v>0 \,.
\end{cases}
\eea

 The minimal value $r_{m}$ of the $r$-coordinate on the BTZ geodesic is
 \begin{equation}
\label{rmin}
r_{m}^{2} = \frac{r_{h}^{2}}{2} \left( 1-E^{2}+J^{2} + \sqrt{\left( 1-E^{2}+J^{2} \right)^{2} -4J^{2}} \right) \, ,
\end{equation}
where $E,J$ are given by eq. (\ref{motion-constants}).

Since the shell of null dust is at $v=0$, the time dependence of the junction point $r_{s}$
 can be determined by imposing that $v(r_{s})=0$ in eq. \Eq{geo2}:
\begin{equation}
\label{tt-eq}
\frac{r_{s}}{r_{h}}= \frac{1}{2} \left( \coth (r_h t)+ \sqrt{ \coth^2 (r_h t) 
- \frac{2 \sqrt{1-\frac{r_{*}^{2}}{r_{s}^{2}}}}{1+ \sqrt{1-\frac{r_{*}^{2}}{r_{s}^{2}}}} } \, \right) \, .
\end{equation}
The system of Eqs. \Eq{tt-eq} and \Eq{ll-eq} 
determine the time dependence of $r_{s}$ and $r_{*}$; unfortunately 
they cannot be solved in closed form.

\begin{figure}[h]
\centering
\begin{tabular}{cc}
\includegraphics[width=7cm]{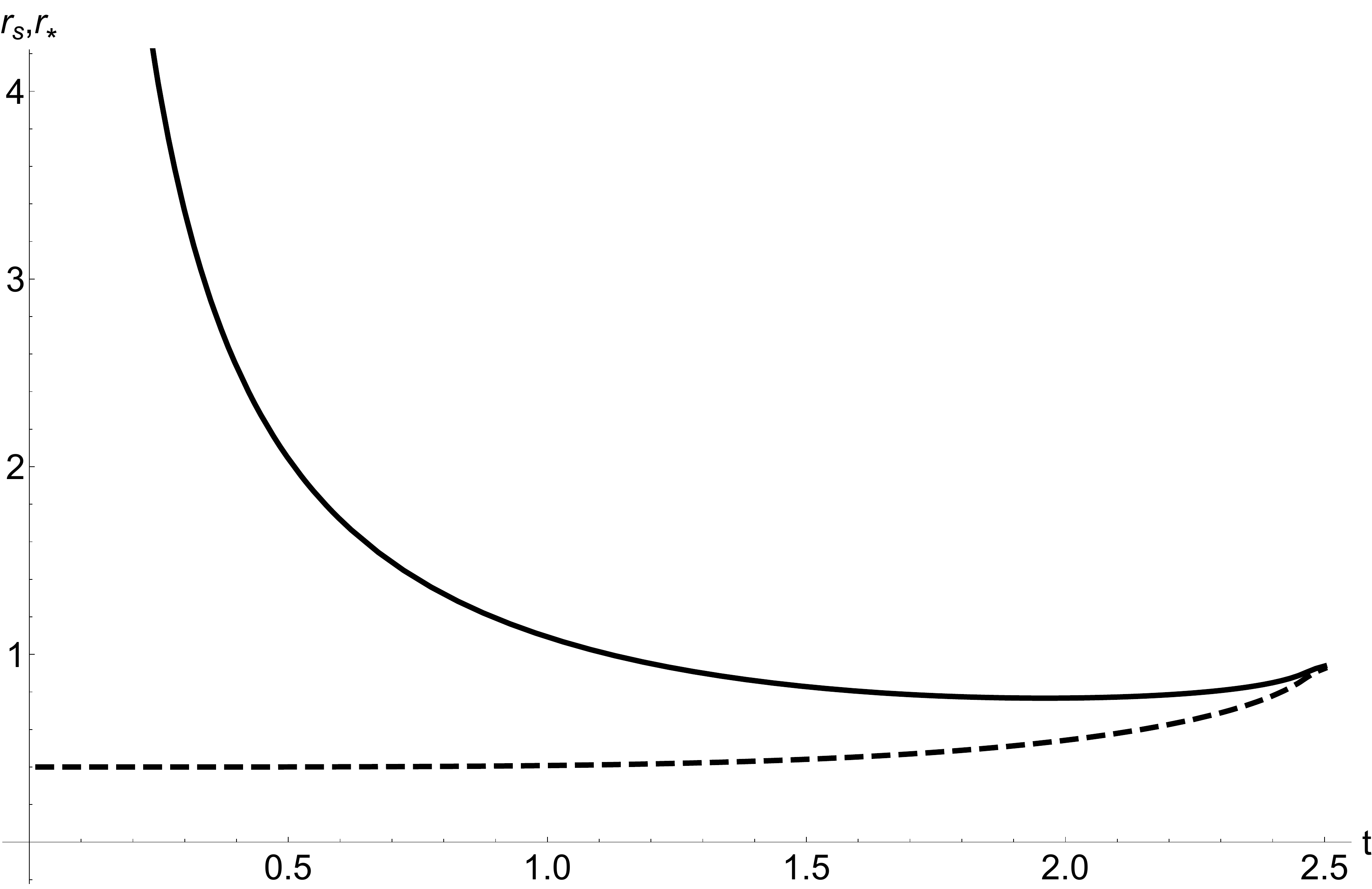} & \includegraphics[width=7cm]{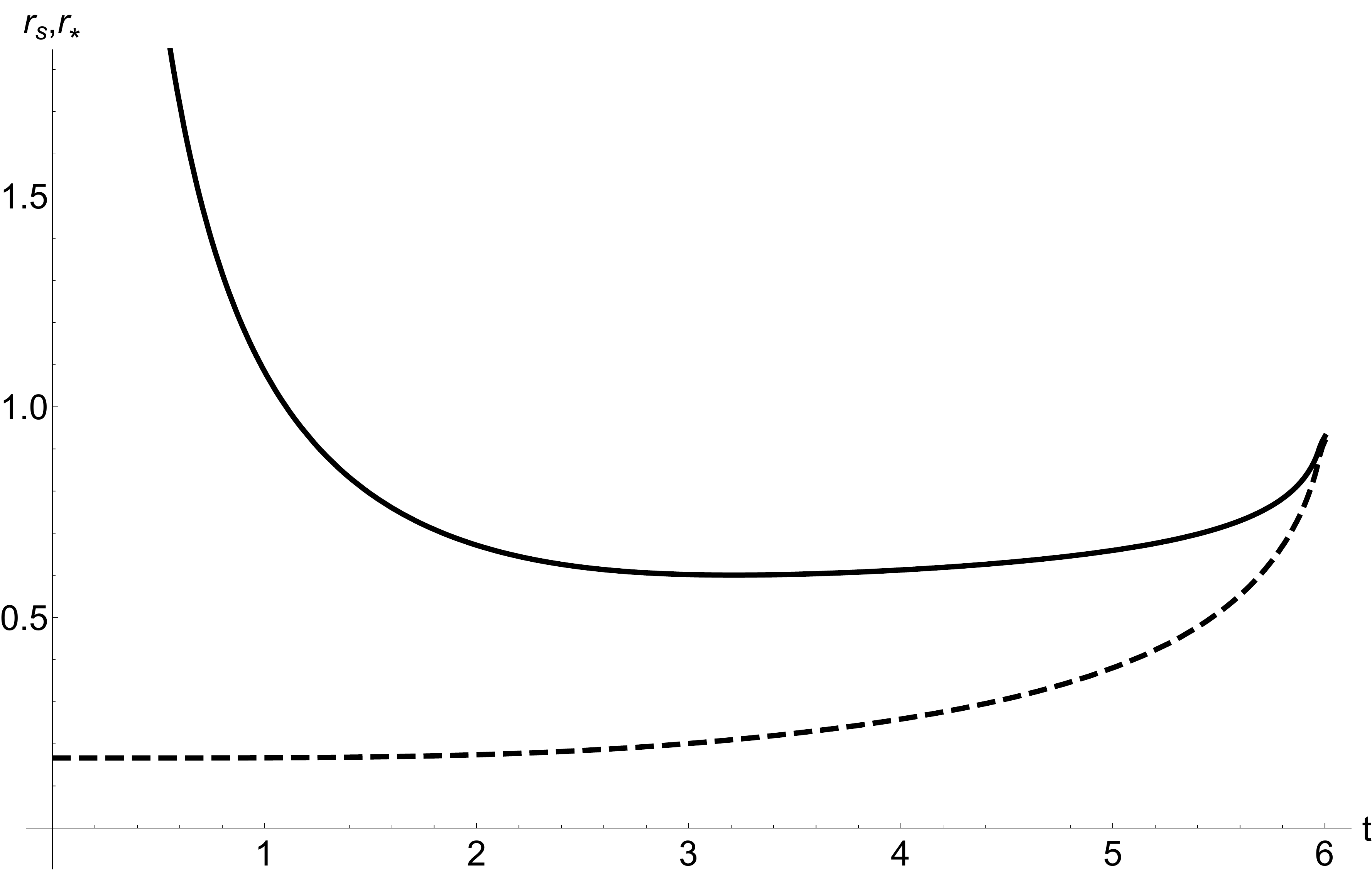}
\end{tabular}
\caption{The plots  show  $r_{s}$ (solid line) and $r_{*}$ (dashed line) as a function of the boundary time $t$.
Here $r_h=1$, and we set $l=5$ on the left and $l=12$ on the right. }
\label{fig:fig5}
\end{figure}

In Fig. \ref{fig:fig5} we show numerical results for particular 
values of the boundary subregion size.  
At $t=0$ the HRT surface entirely 
lies in the AdS part of the full spacetime, and so $r_{s} \rightarrow \infty$ 
and $r_{*}(0) = 2/l$. 
The thermalization time $t_{*}$ is given by the value of the boundary time at which $r_{s}$ and $r_*$ intersect.
For $t>t_*$ the HRT surface entirely lies in the BTZ part of the dynamical spacetime;
from this time the subregion complexity drops to the constant thermal value.
Eqs.  \Eq{tt-eq} and  \Eq{ll-eq} give
 \beq
 t_*=l/2 \, , \qquad
 r_*(t_*)=r_{s}(t_*)=r_{0} \, ,
 \eeq
  see \Eq{coth}.
 For $l \gg 1/r_h$, we have $r_0 \rightarrow r_h$.

\begin{figure}[h]
\center
\includegraphics[width=10cm]{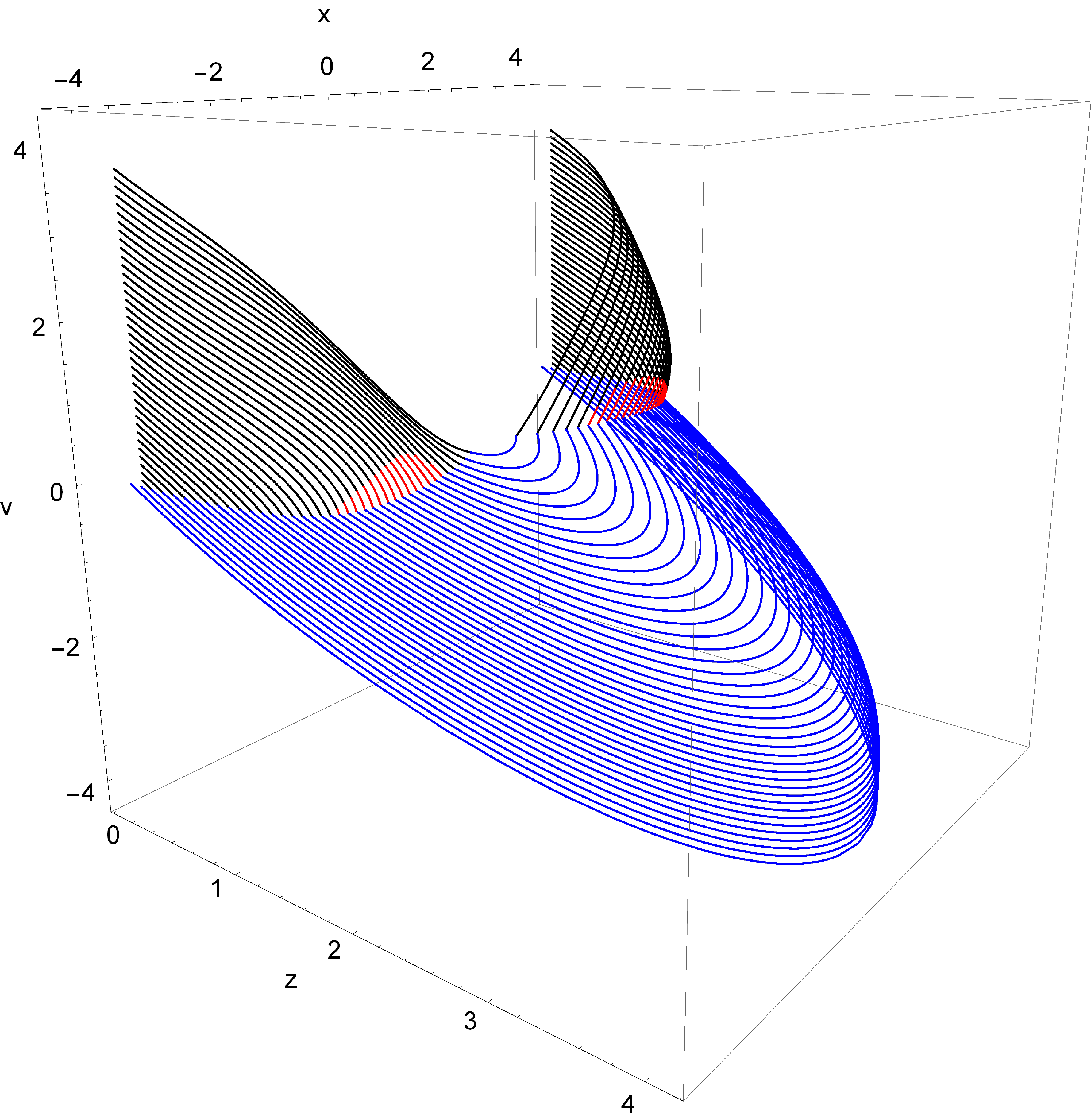}
\caption{Time evolution of the geodesic for $l=8, r_h=1$. The black and red curves respectively denote branch 1 and 2 in the BTZ part; 
the blue curves denote the AdS part of the full geodesic.}
\label{elle8}
\end{figure}

 An example of the time evolution of the geodesics 
is shown in Fig. \ref{elle8}. 

\subsection{Numerical geodesics}

In order to solve the partial differential equations
for the extremal volume, it is useful to 
consider the case of non zero $\tilde{v}$
in eq. (\ref{mass-shell}) in order to make the numerical
problem more tractable.  For generic $\tilde{v}$, one has to solve the geodesics equations numerically:
\bea
&& \ddot{v}+\frac{\dot{v}^2}{z}-\frac{\dot{x}^2}{z}=0 \, ,
\qquad
\ddot{x}-2 \frac{\dot{x} \dot{z}}{z} = 0 \, ,
\nl
&&\ddot{z}+\le z \, m(v)-\frac{2+ z^3 m'(v)}{2 z} \ri \dot{v}^2 -\frac{2}{z} \dot{z}^2 -\frac{2}{z} \dot{z} \dot{v}
+\le \frac{1}{z} - z m(v)\ri \dot{x}^2 = 0 \, ,
\label{geo-num}
\eea
where the dot denotes a derivative with respect to the affine parameter $\lambda$ 
and the $'$ represents a derivative with respect to the coordinate $v$.
The equations are solved with the boundary conditions shown in (\ref{bc1})
 using a shooting method implemented in Mathematica.
In the  $\tilde{v} \rightarrow 0$ limit, we recover the analytical solution
in section \ref{sect-joining}.

\section{Volume}
\label{sect:volume}

In this section we compute the extremal volume of the region
delimited by the segment of length $l$ and the HRT surface
as a function of the boundary time $t$.
This volume has been  proposed to be dual
to mixed state complexity in the boundary CFT 
\cite{Alishahiha:2015rta}.

\subsection{Volume for AdS and BTZ}

In the initial stage ($t \leq 0$) the volume of the region of interest
 is entirely in AdS$_3$, while at final time $t \geq l/2$ the volume is 
 entirely in the BTZ geometry. So these cases correspond to the initial
 and final values of the subregion complexity.
 Moreover, the volume is ultraviolet divergent and a natural
  regularization is given by subtracting the initial 
  AdS volume $V_{AdS}$.
  In this case the boundary geodesic is
  \beq
x^2+z^2=(l/2)^2 \, ,
\eeq
 and the extremal volume solution is given by 
\beq
z= t - v \, .
\eeq
Introducing an UV cutoff at $z=1/\Lambda$, the AdS volume is
\beq
V_{AdS}=
2 \int_{1/\Lambda}^{l/2} \frac{\sqrt{(l/2)^2-z^2}}{z^2} dz=l \, \Lambda -\pi \, .
\label{AdS-vol}
\eeq
The volume at the final equilibrium time
turns out to be exactly the same, i.e. 
\beq
V_{BTZ}=V_{AdS} \, .
\label{vol-btz-ads}
\eeq
This non-trivial   property holds only in 
AdS$_3$ and has topological roots:
it can be proved using the 
Gauss-Bonnet theorem \cite{Abt:2017pmf}.

\subsection{Inconsistency of the $x$-independent ansatz}

Let us parameterise the volume by a surface $v(x,r)$ in AdS$_3$ Vaidya spacetime.
The volume functional can be written as:
\beq
V=\int \, dr \, dx \, \mathcal{V} \, , \qquad \mathcal{V}=\sqrt{r^2 (2-r^2 f \p_r v )\p_r v-(\p_x v)^2 } \, ,
\label{volu}
\eeq
where $f$ is a function only of $r,v$, and let us denote
\beq
v_x=\p_x v \, , \qquad  v_r=\p_r v \, . 
\eeq
The Euler-Lagrange equation gives
\beq
\p_x \le \frac{\p \mathcal{V}}{\p v_x} \ri
+\p_r \le \frac{\p \mathcal{V}}{\p v_r} \ri - \frac{\p \mathcal{V}}{\p v}=0 \, .
\label{volu-eq1}
\eeq
Since the functional (\ref{volu}) is invariant by translations in $x$,
it is reasonable to look for solutions 
of eq. (\ref{volu-eq1}) which are $x$-independent, i.e. 
\beq
v_x=0 \, , \qquad \p_x v_r=0 \, .
\label{x-indip}
\eeq
With the ansatz (\ref{x-indip}), and with the choice
$f=f(r)=f_{BTZ}$, the equation of motion (\ref{volu-eq1}) 
reduces to an ordinary differential equation:
\beq
\left(3 r_h^2-6 r^2\right) v'(r)^2+\left(-3 r^2 r_h^2+r_h^4+2 r^4\right) v'(r)^3-r
   v''(r)+2 v'(r)=0 \, ,
   \label{xx-indep}
\eeq
where the $'$ denotes a derivative with respect to the coordinate $r$. 

The extremal surface used to compute the subregion complexity
of a segment must be attached to the HRT surface,
which in our case is a geodesic.
Consequently, in order for the $x$-independent ansatz to be consistent, 
eq. (\ref{xx-indep}) should be satisfied by the 
geodesic in eq. (\ref{geo2}). This is correct only for the $E=0$
case, which corresponds to the geodesic
used to compute subregion complexity in the
 static BTZ solution.
So, in the time-dependent case, the $x$-independent ansatz 
 \cite{Chen:2018mcc} obtained from the HRT surface
 does not give a solution of the extremal volume equation of motion.
The $x$-independent ansatz gives
an approximate solution in some limits, because
it is exact both at initial time $t=0$ and at final time $t=l/2$.

We will refer to the $x$-independent volume configuration $v(r)$
which is attached to the HRT surface in eq. (\ref{full-geo1}-\ref{full-geo2})
as the pseudosolution. Strictly speaking, this configuration
will satisfy the equations of motion (\ref{volu-eq1})
only at initial time $t \leq 0$ and after thermalization $t \geq l/2$.
We will give numerical evidence that nearby these two regimes
it is a good approximation to the solution of  (\ref{volu-eq1}).

Since the real solution is expected to be a local maximum of the
volume functional, we expect that the volume of the pseudosolution
is lower than the volume of the solution.
We will check this expectation later in some numerical examples.

\subsection{Volume of the pseudosolution}

The total volume  of the pseudosolution $\hat{V}$  is the sum of two contributions:
\begin{equation}
\hat{V}=\hat{V}_{AdS}+\hat{V}_{BTZ} \, .
\label{v-cappello}
\end{equation} 
The  AdS$_{3}$ part  gives
\begin{equation}
\label{ads-volume}
\hat{V}_{AdS}
= -\pi +2 \frac{\sqrt{r_{s}^{2}-r_{*}^{2}}}{r_{*}} +2 \arcsin \frac{r_*}{r_s} \, .
\end{equation}

In the case $v>0$, the surface is given by eq. \Eq{geo2},
in which we must consider the $+$ sign if we are dealing with branch 1 and the $-$ one if we are dealing with branch 2;
 anyway, the choice of the sign does not modify the result for the induced metric determinant $h$
\beq
\sqrt{h}= r \sqrt{\frac{\left( r-J r_{h} \right) \left( r+J r_{h} \right)}{r^{4}
+\left( -1+E^{2}-J^{2} \right) r_{h}^{2} \, r^{2}+ J^{2} \, r_{h}^{4}}} \, ,
\eeq
where the values of  $E$ and $J$ are given by eq. \Eq{motion-constants}.
Therefore, considering the previous discussion about the BTZ portion of the full geodesic, the BTZ part of the volume is given by:
\bea
\label{btz-volume}
\hat{V}_{BTZ}&=& 2 \, \theta \left( r_{s}-\frac{r_{h}}{\sqrt{2}} \right) \int_{r_{s}}^{\Lambda} dr
 \,  \sqrt{h}
 \int_{0}^{\chi^{+}(r)} dx  \nl
&+& 2 \, \theta \left( \frac{r_{h}}{\sqrt{2}}-r_{s} \right) 
\left\{
 \int_{r_{m}}^{\Lambda} dr \, \sqrt{h}
 \int_{0}^{\chi^{+}(r)} dx  + 
 \int_{r_{m}}^{r_{s}} dr \, 
\sqrt{h} 
\int_{0}^{\chi^{-}(r)} dx 
\right\} \, ,
\eea
in which $\Lambda$ is the UV cutoff in the $r$ coordinate.

From eq. (\ref{v-cappello}) we find the following closed form for
the volume of the pseudosolution:
\bea
\label{Vtotal}
\hat{V} &=& -\pi +2 \frac{\sqrt{r_{s}^{2}-r_{*}^{2}}}{r_{*}} +
2 \arcsin \frac{r_*}{r_s} 
\nl
&+&  \int_{r_{s}}^{\Lambda} dr \, \psi(r)
\left[ \frac{1}{r_{h}} \ln \frac{r^{2}-r_{*} \, r_{h} + \sqrt{r^{4}+\left[ -1+\frac{r_{h}^{2} \left( r_{s}^{2}-r_{*}^{2} \right) }{4r_{s}^{4}}-\frac{r_{*}^{2}}{r_{h}^{2}} \right] r_{h}^{2} \, r^{2}+ r_{*}^{2} \, r_{h}^{2}}}{r^{2}+r_{*} \, r_{h} + \sqrt{r^{4}+\left[ -1+\frac{r_{h}^{2} \left( r_{s}^{2}-r_{*}^{2} \right) }{4r_{s}^{4}}-\frac{r_{*}^{2}}{r_{h}^{2}} \right] r_{h}^{2} \, r^{2}+ r_{*}^{2} \, r_{h}^{2}}} + l \right] 
\nl
&+&  \, \theta \left( \frac{r_{h}}{\sqrt{2}} - r_{s} \right)  \left[ \frac{1}{r_{h}} \ln \frac{1-\frac{r_{h}^{2} \left( r_{s}^{2}-r_{*}^{2} \right) }{4r_{s}^{4}}+\frac{r_{*}^{2}}{r_{h}^{2}} - 2 \frac{r_{*}}{r_{h}}}{1-\frac{r_{h}^{2} \left( r_{s}^{2}-r_{*}^{2} \right) }{4r_{s}^{4}}+\frac{r_{*}^{2}}{r_{h}^{2}} + 2 \frac{r_{*}}{r_{h}}} + 2 \, l \right] 
\int_{r_{m}}^{r_{s}} dr \, 
\psi(r)
 \, ,
\eea
where
\beq
\psi(r)= r \sqrt{\frac{\left( r-r_{*} \right) \left( r+r_{*} \right)}{r^{4}+\left[ -1+\frac{r_{h}^{2} 
\left( r_{s}^{2}-r_{*}^{2} \right) }{4r_{s}^{4}}-\frac{r_{*}^{2}}{r_{h}^{2}} \right] r_{h}^{2} \, r^{2}+ r_{*}^{2} \, r_{h}^{2}}} \, .
\eeq

\subsection{Numerical solution}

We would now like to compute the volume of the extremal surface stretching 
inside the region delimited by the HRT surface. 
For convenience, we parameterize\footnote{Indeed,
the solution expressed as $v(z,x)$ is not a single-valued function nearby the 
regions where branch 1 is attached to branch 2. This is not convenient
for numerical calculations.} the extremal surface
through $z(x,v)$, since we expect this function to be be single-valued.
The volume functional is
\beq
V=\int \, dv \, dx \, \mathcal{V} \, , \qquad \mathcal{V}=
\frac{\sqrt{-(2 \p_v z + f(v,z)  )-(\p_x z)^2} }{z^2}\, ,
\label{volume}
\eeq
and denoting
\beq
z_x=\p_x z \, , \qquad  z_v=\p_v z
\eeq
the Euler-Lagrange equations are
\beq
\p_x \le \frac{\p \mathcal{V}}{\p z_x} \ri
+\p_v \le \frac{\p \mathcal{V}}{\p z_v} \ri - \frac{\p \mathcal{V}}{\p z}=0 \, .
\label{volu-eq}
\eeq
More explicitly, the equation for the extremal solution is
\bea\label{eqqq}
- z_{vv} 
+ z_{xx}  ( 2  z_v  +  f)  
-2 z_{vx} z_x
+  (z_x)^2  \frac{(2 f  -z \p_z f + 2 z_v)}{z} 
+4  \frac{ (z_v)^2}{z} && \nl
+3 z_v \frac{(4 f -z \p_z f)}{2 z}
+2 \frac{f^2}{z} - \frac12 f \p_z f - \frac12 \p_v f &=& 0 \, ,
\eea
with the boundary condition specified by the HRT surface.

We solved this equation numerically using both the analytical and numerical geodesics found in Sec.~\ref{sect:geo}, checking that all results match when $\tilde{v}$ is small enough that the numerical solution of eqs. \eqref{geo-num} gives a good approximation to the analytical solution in the $\tilde{v} \rightarrow 0$ limit.

We used the finite-element method implemented in Mathematica, to solve the equations in an adaptive triangulation of the HRT surface, the discretization consisting of cells with maximum size $\mathcal{O}(10^{-4})$ in units of $r_h = 1$. We checked that our results are robust by reproducing them independently with a linearized iterative solver working on a regular rectangular grid meshing the HRT surface.

\begin{figure}[h]
\centering
\begin{subfigure}[c]{0.4\linewidth}
\includegraphics[scale=0.25]{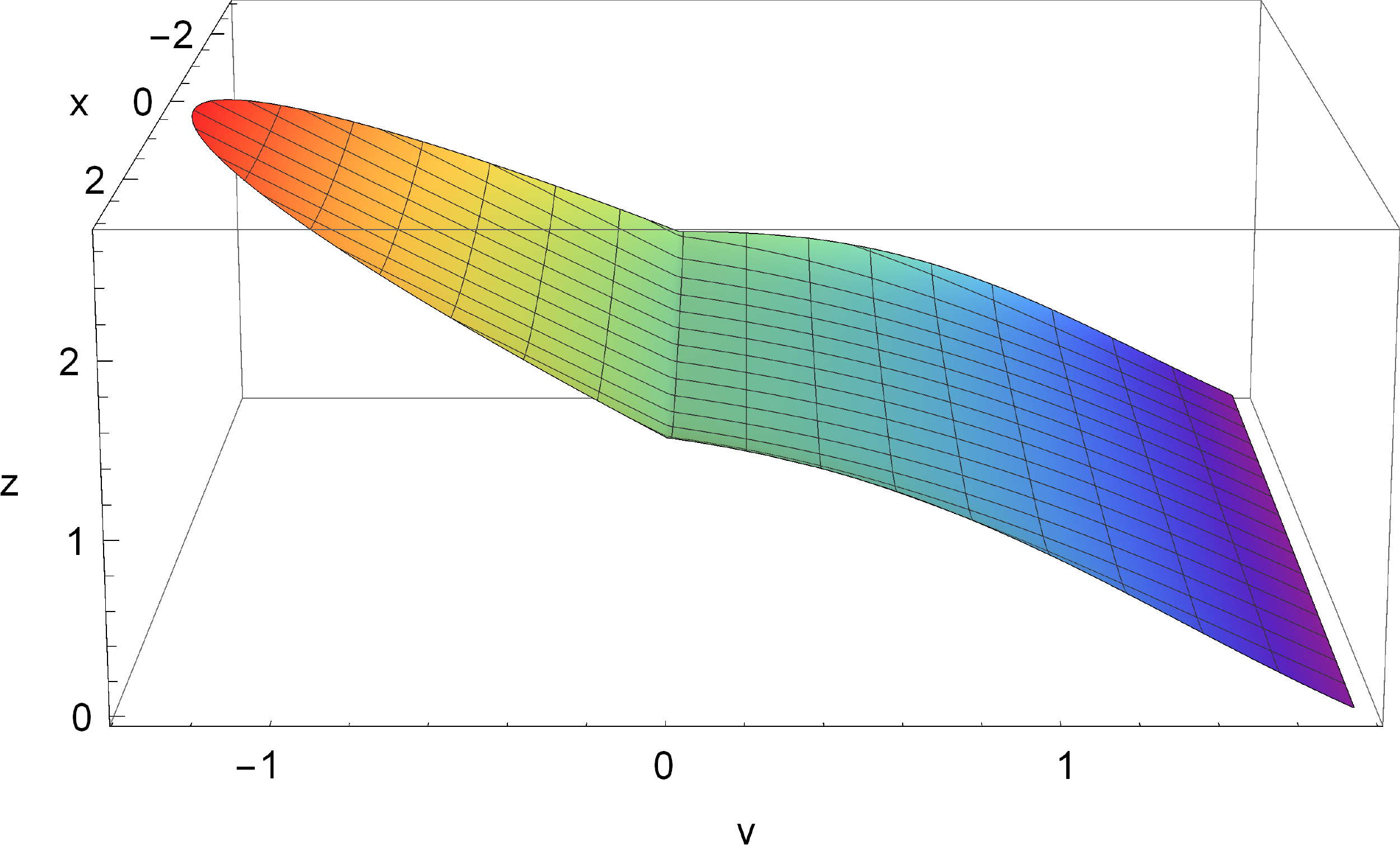}
\end{subfigure} \hspace{15mm}
\begin{subfigure}[c]{0.4\linewidth}
\includegraphics[scale=0.25]{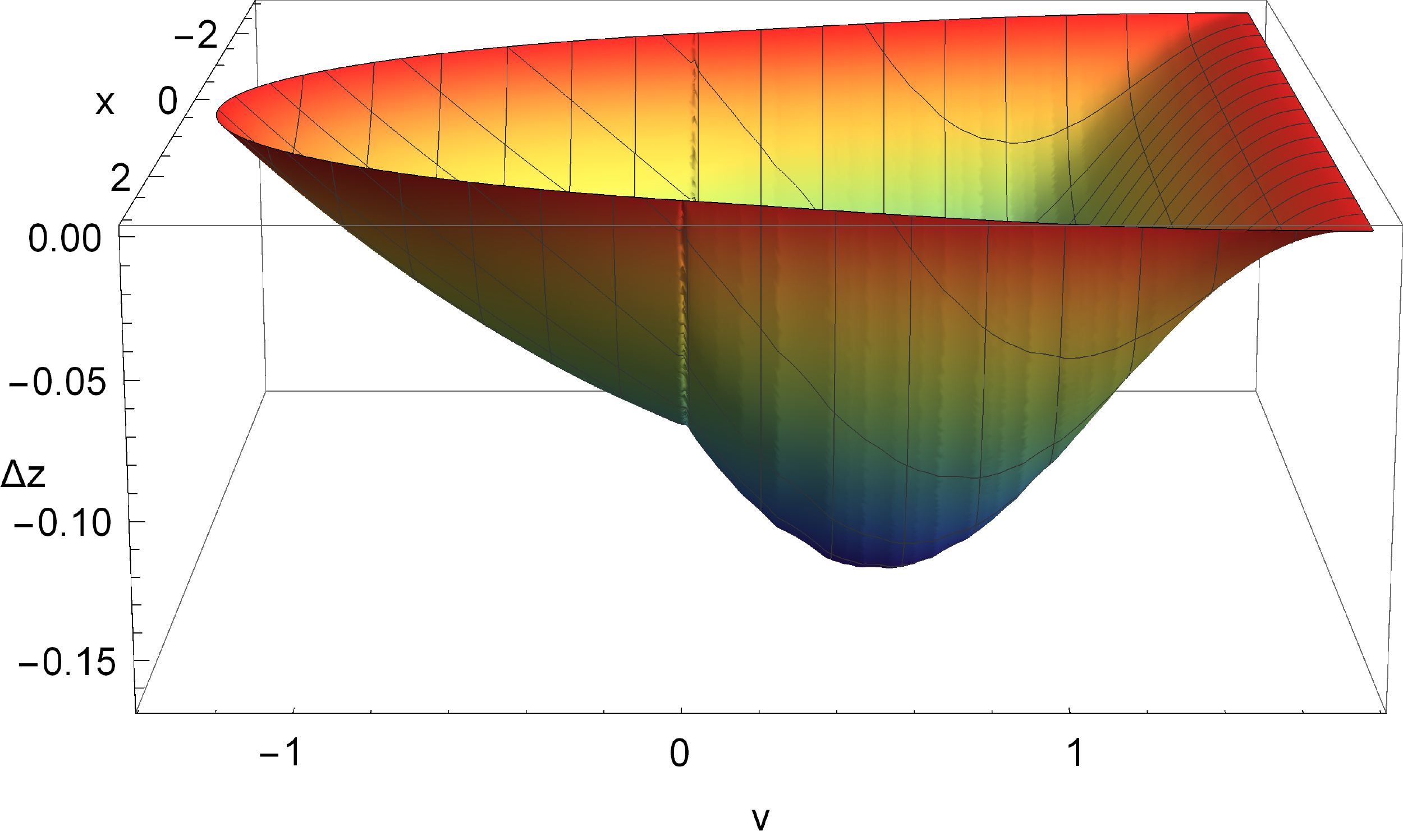}
\end{subfigure}
\caption{Solutions for the extremal volume,
 for $l = 6$, $r_h=1$ and $t = 1.75$. On the left 
we plot the solution; on the right we plot the difference $\Delta z$ between
the solution and the pseudosolution.}
\label{solutions}
\end{figure}

We solved the volume equations numerically up to $r_h l=6$; higher values of $l$ are numerically challenging, 
 because the geodesics develop sharp kinks requiring very fine-grained discretizations in order to obtain reliable results.
An example solution is shown in Fig.~\ref{solutions}.
The geodesics forming the boundary of the HRT surface are not smooth, this is expected from the solutions shown
 in Fig.~\ref{elle8}. 
 As can be seen, there are significant differences between the numerical solution and the pseudosolution.

\subsection{Time dependence of volume}

\begin{figure}[h]
\centering
\begin{tabular}{cc}
\includegraphics[width=7cm]{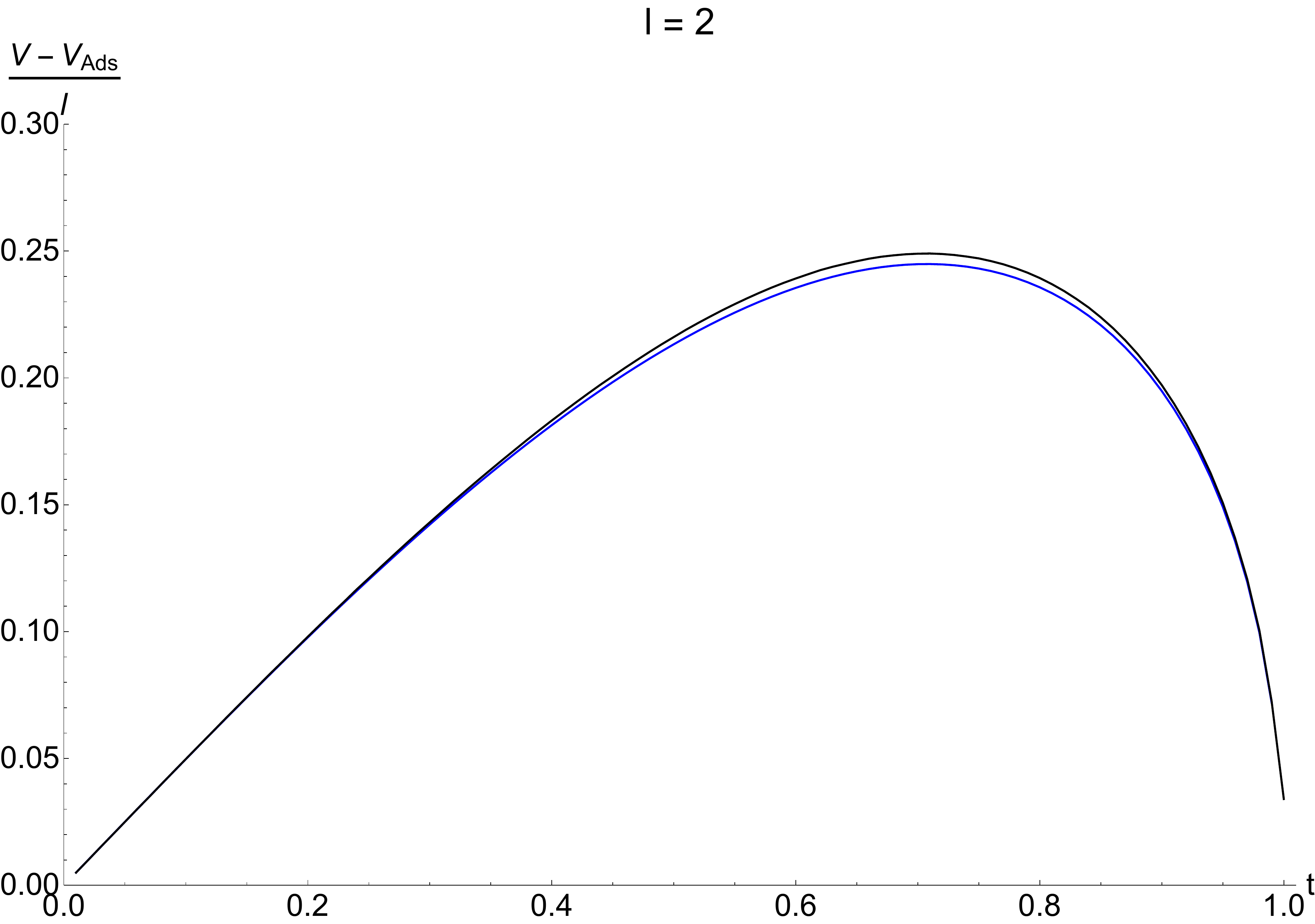} & \includegraphics[width=7cm]{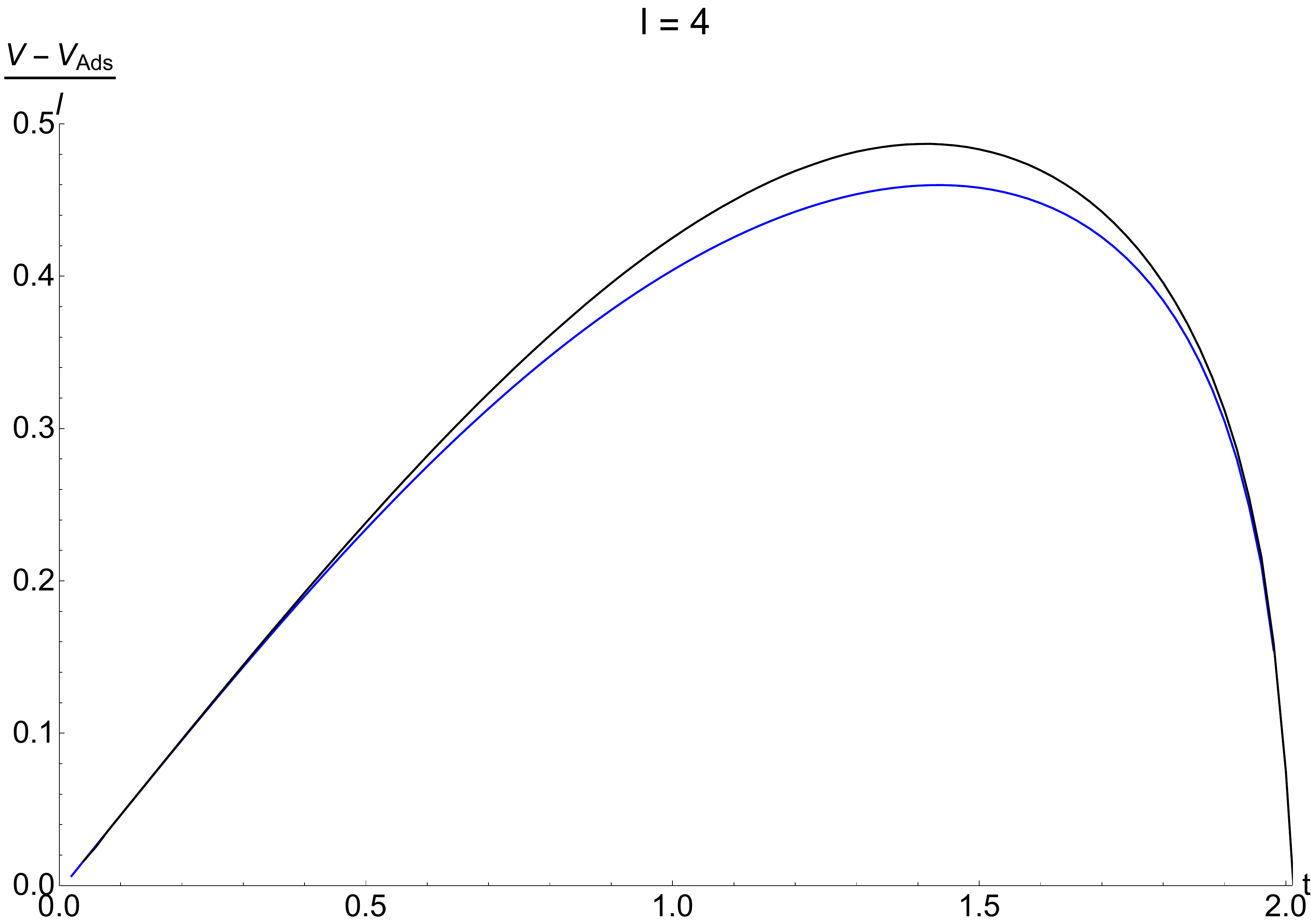}\\
\includegraphics[width=7cm]{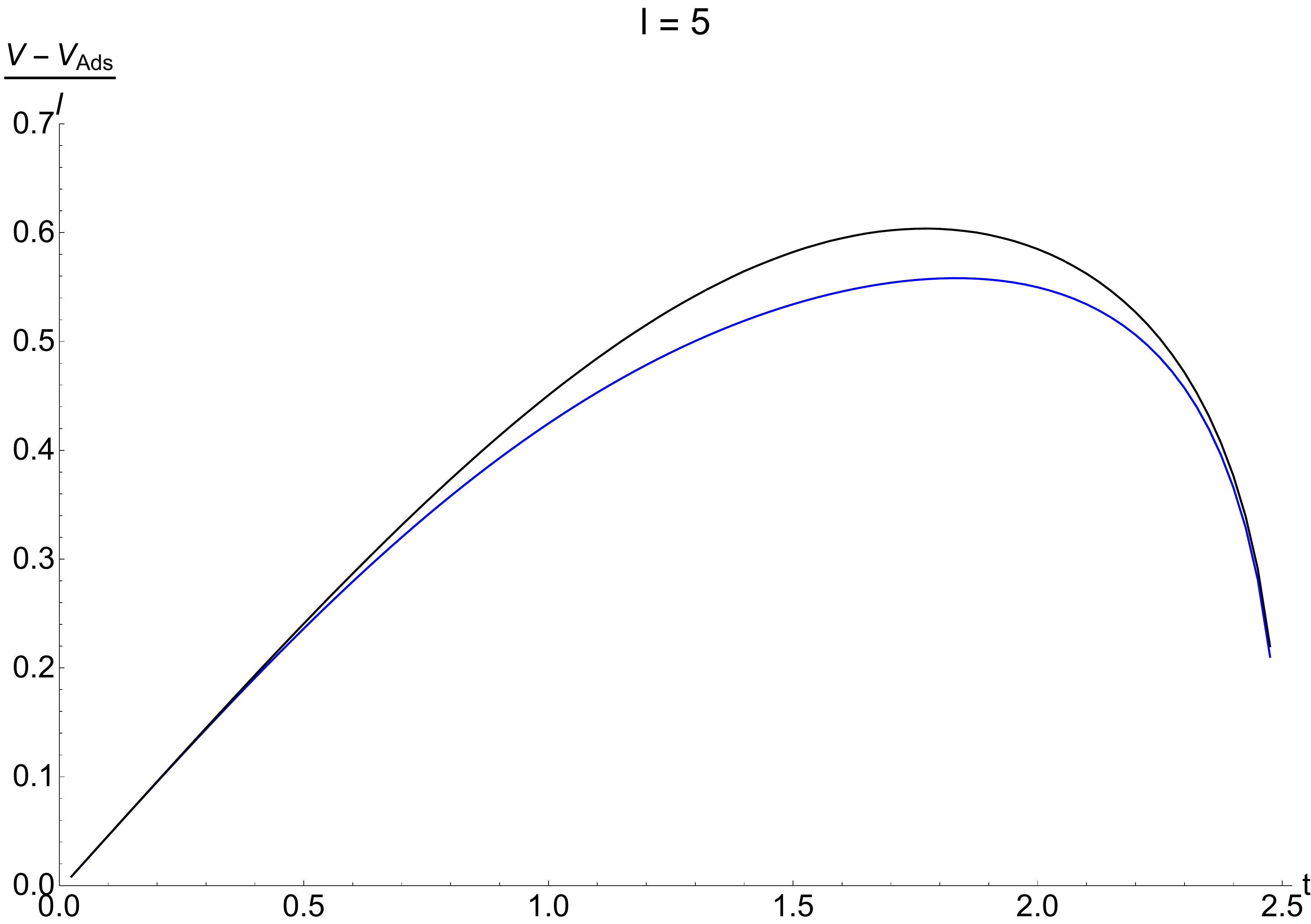} & \includegraphics[width=7cm]{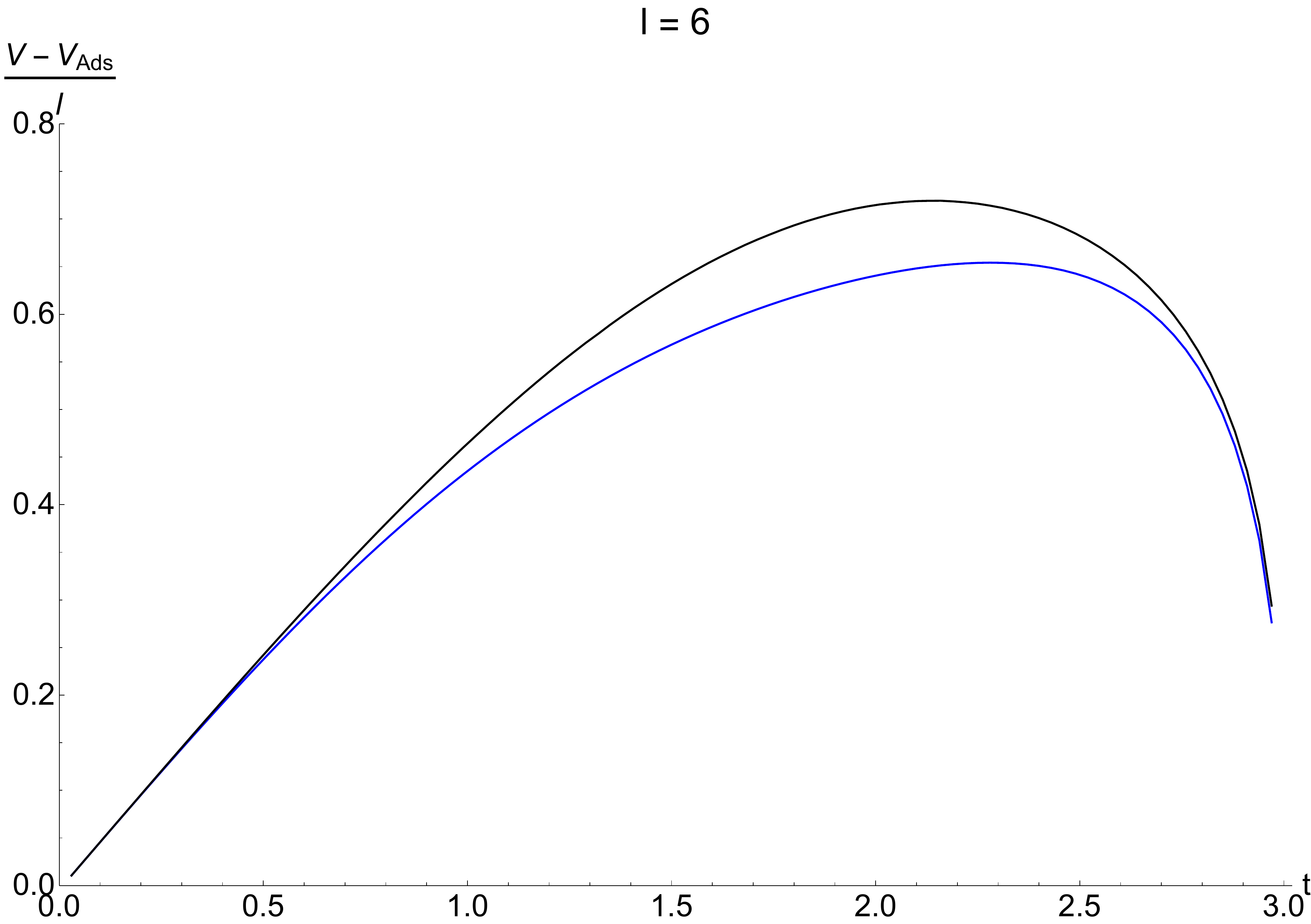}
\end{tabular}
\caption{Time dependence of the volume $V$
of the solution (black), compared to the volume $\hat{V}$ of the pseudosolution (blue)
 for $l=2$ (up, left),
 $l=4$ (up, right),
$l=5$ (left, bottom), $l=6$ (right, bottom).  
 We set $r_h=1$. }
\label{volume-2-4-5-6}
\end{figure}

We are then interested in the volume functional
(\ref{volume})
 evaluated on the equation of motion, which we denote by $V$.
 We regularize UV divergences by 
subtracting the AdS volume (\ref{AdS-vol}).
The volume of the solution as a function of the time $t$ is 
shown in Fig.~\ref{volume-2-4-5-6};
for comparison, also the  volume of the pseudosolution
is displayed.  
The solution has indeed as expected a bigger volume.
Fig.~\ref{volume-2-4-5-6} 
confirms that the volume of the pseudosolution
is indeed a good approximation both for early $t \approx 0$
and late $t \approx l/2$ times.
For intermediate times,
the discrepancy tends to increase with $l$.
 As can be seen, the plot of the volume of the numerical solution seems 
to be smoother than the one of the pseudosolution. In particular,
the variation of the slope of the solution is less pronounced
than the one of the pseudosolution.

\subsection{Analytical results}

Both at early times and at late times, the volume of the pseudosolution
is a good approximation of the volume of the solution.
It should be remarked that the pseudosolution in any case provides
a lower bound of the volume of the solution.

When $l$ is large enough, typically larger than $1/r_h$,
there are three stages in the evolution of the volume of the pseudosolution:
\begin{itemize}
\item Early times.
If we replace the early time results eq. (\ref{early-bird})
in the volume expression eq. (\ref{Vtotal}),
 we find, at the leading order in $l$:
\beq
\frac{\hat{V}}{l}=\Lambda+ r_h \, \tanh \frac{r_h t}{2} + \mathcal{O} (1/l)\, .
\label{early-volume}
\eeq
This is true in both the regimes $r_s >r_h/\sqrt{2}$
and $r_s < r_h/\sqrt{2}$; the only assumption is 
that time is so early that eq. (\ref{early-bird}) can be trusted.
From numerical evidence, it turns out that this part of the evolution continues 
for a time that scales as $  \mathcal{O} (\log (r_h l))$.

At early times, the pseudosolution is a good approximation
to the full solution. In particular, one can safely trust
the first order Taylor expansion of eq. (\ref{early-volume}), i.e.
\beq
\frac{{V}}{l}=\Lambda+   \frac{r_h^2 t}{2} +  \mathcal{O} (1/l) \, .
\label{tanacca}
\eeq
This is further supported by the fact that $\tanh x \leq x$ and that 
the volume of the pseudosolution is a lower bound of the one of the solution.
This agrees with the result in eq. (3.77) of
 \cite{Chapman:2018dem} for the growth rate 
 $\tilde{V}$ of the volume in a one-sided Vaidya black hole,
which in our notation and for $d=2$ reads:
\beq
\frac{d \tilde{V} }{dt} =\Omega_{k} \frac{r_{h}^{2}}{2 } \, ,
\label{pure-c}
\eeq
where $\Omega_{k}$ is divergent and it corresponds 
to our boundary subregion size $l$ in the limit $l \rightarrow \infty$.

\item Intermediate times, $\mathcal{O} (\log r_h l ) < t < \frac{l}{2}-\frac{0.53}{r_h}$.
An explicit analytical formula for the volume
 of the pseudosolution at large $l$ is derived in appendix \ref{appe-late-time}:
\beq
\frac{\hat{V}}{l} \approx \Lambda + \frac{\mathcal{I}_1}{l} + (\Upsilon-1) \eta(r_s) - \Upsilon \eta(r_m) \, .
\label{tempi-medi}
\eeq
where  ${\mathcal{I}_1}$, $\Upsilon$, $\eta$ are defined in Appendix  \ref{appe-late-time}.
Unfortunately, at large $l$ we expect significant deviations
between the solution and pseudosolution volumes.
Nonetheless, this estimate is still useful
because it provides a lower bound to the volume of the solution.

\item Late times, $\frac{l}{2}-\frac{0.53}{r_h}<t<\frac{l}{2}$.
We can approximate the volume of the pseudosolution as
\beq
\frac{\hat{V}}{l} 
= \Lambda+
\frac{r_h^2 l}{4} \frac{\sqrt{r_s r_h (r_h-r_s) (2 r_s-r_h ) }}{(r_s-r_h)^2 +r_s^2 } +\mathcal{O}(l^0) \, ,
\label{tempi-lunghi}
\eeq
see appendix \ref{appe-late-time} for a derivation.
The maximum of $\hat{V}$ is at $r_s=r_h/\sqrt{2}$ and scales as:
\beq
{\rm max} \, \le \frac{\hat{V}}{l}  \ri  = \Lambda +\frac{l r_h^2}{8}  \, .
\label{massi}
\eeq
Using the approximation eq. (\ref{time-late-time}), we find
the following behaviour nearby $t \approx l/2$:
\beq
r_s \approx r_h \left(1 -\sqrt{\frac{r_h}{2} \left( \frac{l}{2} -t \right)}\right) \, , \qquad
\frac{\hat{V}}{l} \approx \Lambda +l \left(\frac{r_h}{2} \right)^{9/4} \left( \frac{l}{2} -t\right)^{1/4} \, .
\label{tempi-lunghi-2}
\eeq
Also in this regime we expect that this
is a good approximation of the volume of the solution.

\end{itemize}

\subsection{Discussion}

\begin{figure}[h]
\center
\includegraphics[width=10cm]{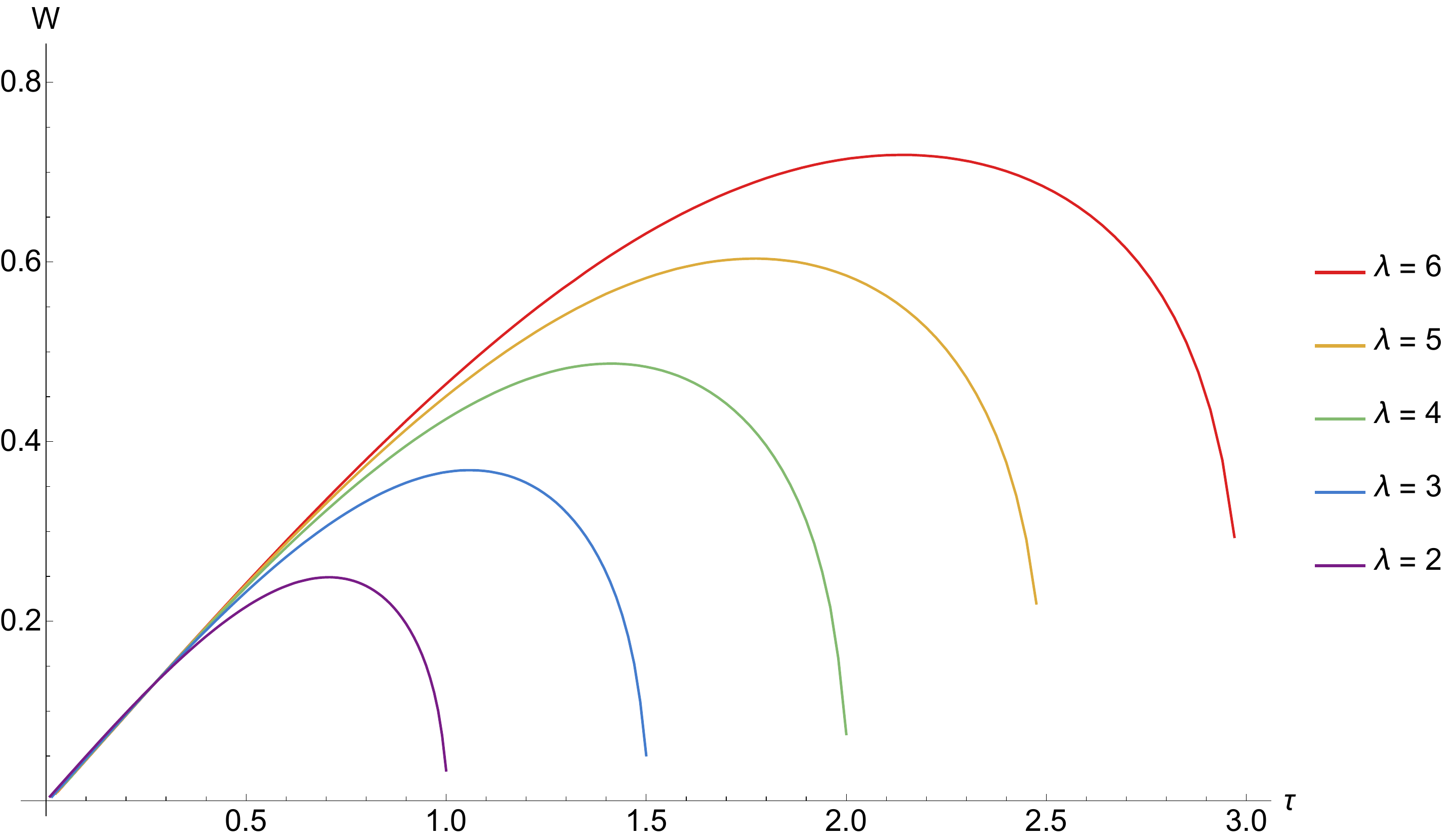}
\caption{$W_\lambda $ as a function of $\tau$ for some values of $\lambda$.}
\label{many-LL}
\end{figure}

The central charge of the boundary theory $c$, the final temperature $T$, entropy $S$
and  complexity $\mathcal{C}_V$ 
can be expressed in terms of bulk quantities as follows
\beq
c=\frac{3}{2 G }  \, , \qquad T=\frac{r_h}{2 \pi} \, , \qquad
S=\frac{r_h l }{4 G} \, , \qquad
\mathcal{C}_V=\frac{V}{G} \, ,
\eeq
where we set the AdS radius $L_{AdS}=1$ by a choice of units.
The regularized complexity, defined as $\Delta \mathcal{C}_V =  \mathcal{C}_V - \mathcal{C}_V^{AdS}$,
can be expressed as 
\beq
\frac{\Delta \mathcal{C}_V}{l} =  \frac{4 \pi}{3} \, c  \,  T  \,  W_\lambda (\tau) \, , 
\eeq
where
\beq
\tau=2 \pi T \,  t \,  ,  \qquad \lambda= 2 \pi T \, l \, , \qquad W_\lambda (\tau)=\frac{V_{\rm sol}-V_{\rm AdS}}{\l} \, .
\eeq
The function $W_\lambda(\tau)$ is plotted in Fig.~\ref{many-LL} for a few values of $\lambda$.
For small $\tau$,  from eq. (\ref{tanacca}) we find $W_\lambda \approx \tau/2$.

\begin{figure}[h]
\center
\includegraphics[width=10cm]{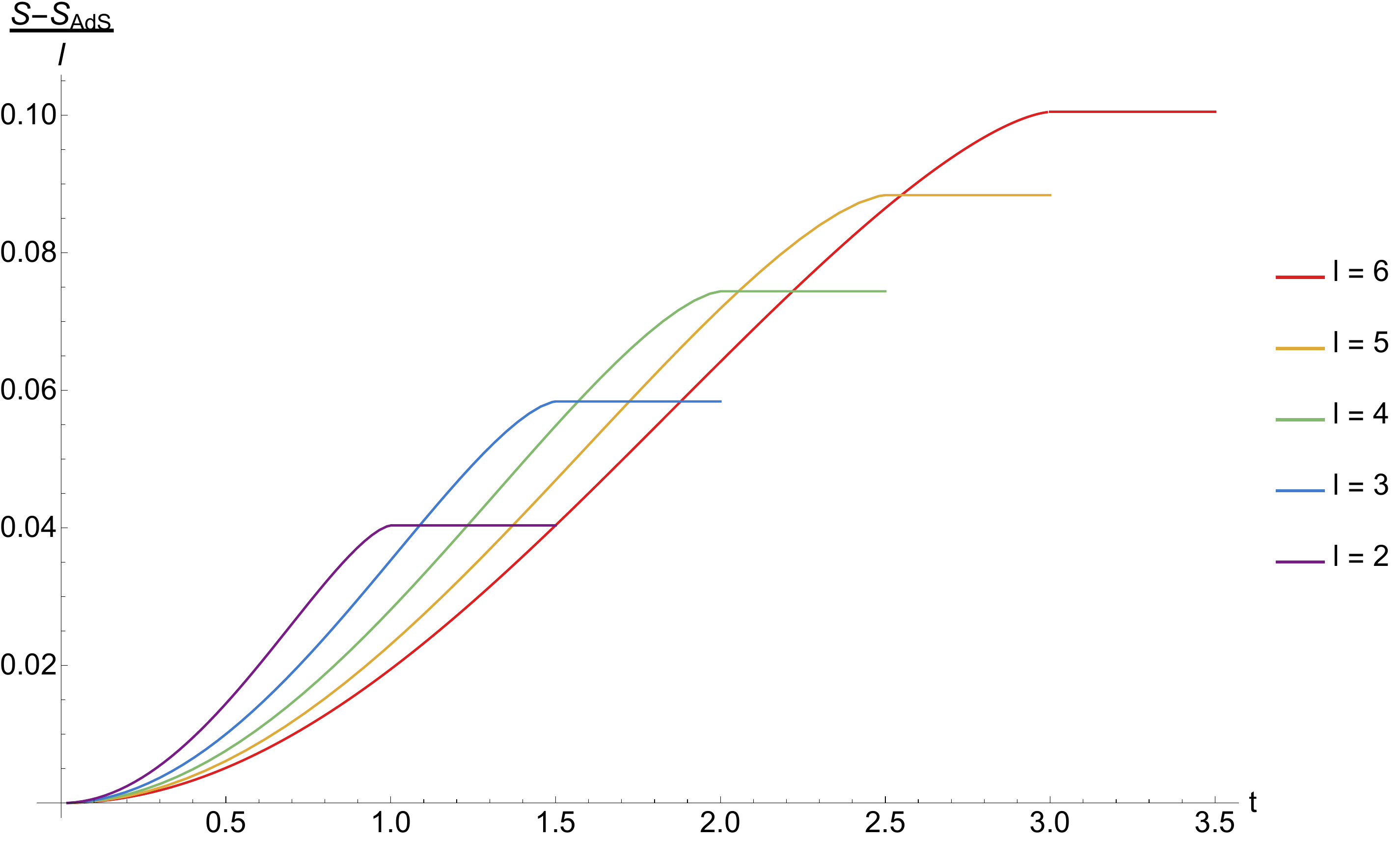}
\caption{Entropy  of the solution as a function of time for some values of $l$,
with the diverging entropy of the vacuum AdS subtracted.
Here we set $G=1$, $r_h=1$ for illustrative purposes.}
\label{entropy-fig}
\end{figure}

It is interesting to compare the time behaviour of complexity
with the one of entanglement entropy,
which can be computed using
eq. (109) of  \cite{Balasubramanian:2011ur}.
A plot is shown in Fig \ref{entropy-fig}.
While the behaviour of entanglement entropy 
interpolates between the value in AdS and the thermal one
in a monotonic way during the quench, the behaviour of subregion complexity
grows to a maximum which scales as $l^2$ and then goes back to the original
value of empty AdS.

It is remarkable that, after thermalization, $\Delta \mathcal{C}_V=0$,
eq. (\ref{vol-btz-ads}). From the geometrical point of view, this property
follows from the Gauss-Bonnet theorem. From the point of view of the boundary
field theory this behaviour looks rather counterintuitive. 
Indeed, for asymptotically AdS$_d$ black branes,  with $d>3$, this
property does not hold \cite{Ben-Ami:2016qex}.
On the other hand,  in the small $l \,T$ regime,
the calculations for $d>3$ in \cite{Chen:2018mcc} should be correct. Then
we expect that, also in this case, subregion complexity, 
after the initial growth stage, decreases
at large times going back to a value which is much closer to the
original one compared to its maximum.

We can qualitatively interpret this behaviour as follows.
One of the most promising candidates for the field theory dual
of subregion complexity is purification complexity, 
 which is defined as the minimal pure state complexity among all possible purifications
of the given mixed state \cite{Agon:2018zso}.
At equilibrium, there is a maximal amount of  possible pure microstates 
which corresponds to the given mixed macrostate.
In this big community of states, it should not be surprising that the minimal
complexity is small, due to the large number of samples.
 Instead, far away from equilibrium, the number of microstates
which describe our density matrix is much smaller,
 and so we can expect that the minimal
complexity is bigger.

 We expect that the  Lloyd's bound \cite{Lloyd} should apply only 
 when subregion complexity coincides with the pure state one.
 This should be true only at early times, because the boundary effects are negligible.
 Indeed  in this regime $W_\lambda \approx \tau/2$ and then
 we recover the result (\ref{pure-c}):
 \beq
 \frac{d{\mathcal C}}{dt}= 8 \pi M \, , \qquad M=\frac{l \, r_h^2}{16 \pi G} \, ,
 \eeq
 where $M$ is the black hole mass.
This is the same as the asymptotic complexity rate
in time-independent black holes, and as such saturates
the conjectured Lloyd's bound. Moreover from figure \ref{many-LL}
we see that, nearby $t=0$, the rate  $\frac{d{\mathcal C}}{dt}$
is a decreasing function of time, and so the Lloyd bound is not violated
also by subregion complexity at small time. 

\section{Conclusions}
\label{sect:conclu}

In this paper we studied the holographic subregion volume complexity 
for a line segment of length $l$ in the AdS$_3$ Vaidya geometry, in the limit of zero shell thickness
eq. (\ref{heaviside}). We computed the extremal volume as a function of time
numerically, and we found that both at early times $t \approx 0$
and at late times, nearby equilibrium $t \approx l/2$, the 
$x$-independent  ansatz is a good approximation of the solution
for the extremal volume.
We give  analytical expressions for the extremal volume in both the
early and late time regimes, see 
eqs. (\ref{early-volume}) and (\ref{tempi-lunghi},\ref{tempi-lunghi-2}).
In particular, the maximum of the volume  of the
pseudosolution scales as $l^2$, see eq. (\ref{massi}).
Since the pseudosolution is a lower bound of the solution,
we expect that the maximum of the volume of the solution
scales at least as $l^2$.
 
 We were able to numerically study the full dependence of holographic subregion volume
 complexity (see figure \ref{many-LL}) just for  $r_h l \leq 6$.
 Figure \ref{volume-2-4-5-6}
shows that the corrections from the $x$-independent
 pseudosolution become increasingly important as $l$ grows.

Several problems call for further investigation:
\begin{itemize}
\item
 It would be interesting to study
 larger values of $l$, because it is the regime where bigger deviations
 from the $x$-independent pseudosolution are expected.
  In particular, in \cite{Chen:2018mcc,Ling:2019ien}
 it was conjectured that for large $l$ and intermediate times
 a linear increase regime of complexity holds,
 with a different slope compared to the early times regime.
This conjecture was based on the calculation of the volume of
the $x$-independent pseudosolution. However, since we showed that 
at large $l$ one should expect large deviation between the volumes
of the solution and the pseudosolution, this conjecture should be revisited.
\item Another open problem is to study the time evolution of subregion
action complexity during a quench and to compare it to the volume.
In many cases the action and the volume conjectures give qualitatively similar results
(there are however some exceptions, see e.g. \cite{Chapman:2018bqj}),
which makes hard to discriminate between them.
Due to the large arbitrariness  in several technical aspects of the  definition
of complexity in QFT, it could also be that each of the conjectures is dual to a different 
field theory definition of quantum computational complexity.
\item 
It would be interesting to study complexity evolution
during a quench in QFT. This was initiated in 
\cite{Alves:2018qfv,Camargo:2018eof} for free field theories.
\item There are several possible definitions of subregion complexity in a quantum theory,
for example purification and basis complexity \cite{Agon:2018zso}.
It would be interesting to establish robust properties of these  quantum information
quantities, in order to eventually match them with holographic conjectures.
Another interesting direction is fidelity \cite{Alishahiha:2015rta}.
\end{itemize}

 \section*{Acknowledgments}

G.T is funded by Fondecyt grant 11160010.

\section*{Appendix}
\addtocontents{toc}{\protect\setcounter{tocdepth}{1}}
\appendix

\section{Spacelike geodesics in the BTZ black hole}
\label{Appe-btz-geo}

For completeness, in this appendix we briefly sketch the computation of spacelike
geodesics in the BTZ black hole background, following \cite{Balasubramanian:2011ur}.
Introducing the bulk time $t$, the metric is:
\beq
ds^2=-(r^2-r_h^2) dt^2 + \frac{dr^2}{r^2-r_h^2} + r^2 dx^2 \, .
\eeq
The relation between $t$ and the Eddington-Finkelstein coordinate $v$ which is
used in the main text is:
\beq
t=v-\frac{1}{2 r_h} \log \frac{| r-r_h |}{r+r_h} \, .
\eeq
Parameterizing the geodesic length by $\s$, the geodesic equations are:
\beq
-r_h E =-(r^2-r_h^2) \dot{t} \, , \qquad r_h J = r^2 \dot{x} \, , \qquad
1 = -(r^2 -r_h^2) \dot{t}^2 + \frac{\dot{r}^2}{r^2-r_h^2} + r^2 \dot{x}^2 \, ,
\label{GEO}
\eeq
where dot denotes derivative with respect to $\s$.
The parameters $E$ and $J$ are respectively the constants of motion
associated to $t$ and $x$ translation invariance, i.e. energy and angular momentum.
The equations in (\ref{GEO}) can be solved analytically 
(see \cite{Balasubramanian:2011ur}).
The solutions are expressed as $x(r)$ and $v(r)$
 in eqs. (\ref{geo1}), (\ref{geo2}).
The boundary conditions are chosen in such a way that 
the solution  is symmetric under $x \rightarrow -x$.

\section{Analytical approximations for the constraint equations}

\label{appe-constraints}

The constraints in eqs. \Eq{tt-eq} and  \Eq{ll-eq} 
cannot be solved in closed form, and are also rather tricky 
to be solved numerically, due to the exponential accuracy
which is needed at large $l$ and $t$. It is then useful to 
use some approximations which are valid respectively in the early and 
in the late time regime:

\begin{itemize}

\item {\bf Early time approximation}

At early time $r_s \rightarrow \infty$ and $r_* \approx 2/l$, 
so we can use the $r_* \ll r_s$ approximation in 
eqs.  (\ref{tt-eq}) and  (\ref{ll-eq}). This gives:
\beq
r_* = \frac{2}{l} \, , \qquad
 r_{s} = \frac{r_{h}}{2} \coth \left(\frac{ r_{h} t }{2} \right)  \, ,
 \label{early-bird}
\eeq
which provides a good description of the early evolution of the geodesic.

\item {\bf Late time approximation}

If we formally set $t\rightarrow \infty$ in eq. (\ref{tt-eq}), we find the solution:
\beq
\hat{r}_*=r_h \frac{r_s (2 r_s-r_h)}{(r_h-r_s)^2 +r_s^2} \, .
\label{largetime}
\eeq
The curve (\ref{largetime}) is shown in fig. \ref{mappa}, with several $l$-constant  
curves solving the constraint in eq. (\ref{ll-eq}).

\begin{figure}[h]
\center
\includegraphics[width=8cm]{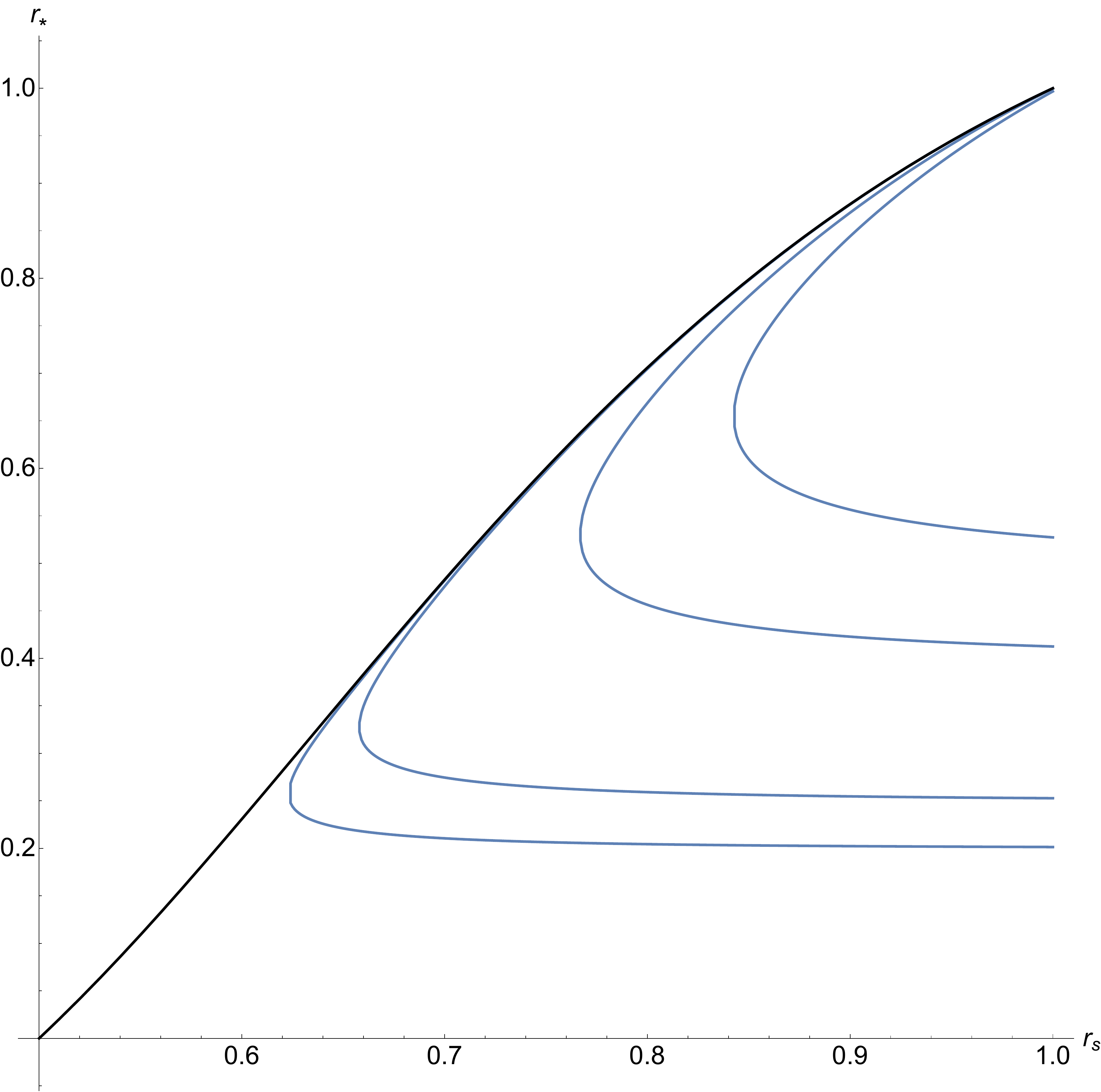}
\caption{ Plot of the $t \rightarrow \infty$ limit curve $\hat{r}_*(r_s)$ (black line), 
with $r_h=1$. The blue lines correspond to $l$-constant  curves
in the $(r_s,r_*)$ plane, see eq (\ref{ll-eq}),  for $l=4,5,8,10$ from top to bottom.}
\label{mappa}
\end{figure}

The physical accessible region of parameters in the $(r_s,r_*)$  plane is below this curve;
as a consequence, we have that $r_s>r_h/2$. 
In the late time regime we can parameterise the deviation from the curve (\ref{largetime}) by 
\beq
r_*=\hat{r}_*- r_h \, \epsilon_*\, ,
\label{espansion-late-time}
\eeq
with a small parameter $\epsilon_* \geq 0$.
We can then solve eq. (\ref{ll-eq}) at the leading order in $\epsilon_*$:
\beq
\label{eppi-appi-2}
\epsilon_*  =  \frac{8(1-\rho_s)(2 \rho_s -1) \rho_s^3}{(1-2 \rho_s +2 \rho_s^2)^2} 
\frac{1}{ \exp \left( r_h l -  \frac{4 (1-\rho_s)}{2 \rho_s -1}  \right) - 
\frac{8 \rho_s^5-20 \rho_s^4+18 \rho_s^3-7 \rho_s+2}{\left(1-\rho_s\right) \left(2 \rho_s^2-2
   \rho_s+1\right)}} \, ,
\eeq
where we have introduced $\rho_s=r_s/r_h$.
Taking the leading large $l$ term we find a simpler expression:
\beq
\label{eppi-appi}
\epsilon_* \approx
\frac{8(1-\rho_s)(2 \rho_s -1) \rho_s^3}{(1-2 \rho_s +2 \rho_s^2)^2}  \exp \left(- r_h l +  \frac{4 (1-\rho_s)}{2 \rho_s -1} \right) \, ,
\eeq
which is a good approximation when $\rho_s$ is not very nearby to $1/2$, 
which is true at large times.

In order to find an approximate expression for $t$ as a function of $\rho_s,\epsilon_*$,
we use then the expansion (\ref{espansion-late-time})
in  the time constraint (\ref{tt-eq}), which gives:
\beq
r_h t=\frac{1}{2} \ln \left(\frac{8 \left(1-\rho_s\right) \rho_s^3}{\epsilon_*  \left(2 \rho_s-1\right) \left(2
   \rho_s^2-2 \rho_s+1\right){}^2}\right) \, .
\eeq
Inserting also the value of $\epsilon_*$ from eq. (\ref{eppi-appi}), we get
\beq
r_h t=\frac{1}{2} \left(r_h l-\frac{4 \left(1-\rho_s\right)}{2 \rho_s-1}-\ln \left( \left(2
   \rho_s-1\right){}^2\right)\right) \, .
   \label{time-late-time}
\eeq
Note that $\rho_s=1/\sqrt{2}$ corresponds to:
\beq
r_h t= \frac{r_h l}{2}-\sqrt{2}-\ln \left(\sqrt{2}-1\right)
\approx \frac{r_h l}{2} -0.53 \, .
\label{tempaccio}
\eeq


\end{itemize}

\section{The volume of the pseudosolution at late time}
\label{appe-late-time}

The approximation in this appendix refer to the limit $l,t \gg 1/r_h$
and to the regime in which $r_*\approx \hat{r}_*$.
We will extensively use  the results of appendix \ref{appe-constraints}.
Let us write the volume of the pseudosolution eq. (\ref{Vtotal}) as:
\beq
\hat{V}=\mathcal{I}_1+\mathcal{I}_2+\mathcal{I}_3+\mathcal{I}_4 \, ,
\eeq
where
\beq
\mathcal{I}_1=-\pi +2 \frac{\sqrt{r_{s}^{2}-r_{*}^{2}}}{r_{*}} +
2 \arcsin \frac{r_*}{r_s} \, , \qquad
\mathcal{I}_2=l \int_{r_s}^\Lambda \psi(r) dr \, ,
\eeq
\beq
\mathcal{I}_3=
\int_{r_{s}}^{\Lambda} \psi(r) \,  \kappa(r) dr \, , \qquad
\mathcal{I}_4= \theta \left( \frac{r_{h}}{\sqrt{2}} - r_{s} \right) l \,  \Upsilon
\int_{r_{m}}^{r_{s}} \psi(r) \, 
dr \, ,
\eeq
where
\bea
\kappa(r)&=&
 \frac{1}{r_{h}} \ln \frac{r^{2}-r_{*} \, r_{h} + \sqrt{r^{4}+\left[ -1+\frac{r_{h}^{2} \left( r_{s}^{2}-r_{*}^{2} \right) }{4r_{s}^{4}}-\frac{r_{*}^{2}}{r_{h}^{2}} \right] r_{h}^{2} \, r^{2}+ r_{*}^{2} \, r_{h}^{2}}}{r^{2}+r_{*} \, r_{h} + \sqrt{r^{4}+\left[ -1+\frac{r_{h}^{2} \left( r_{s}^{2}-r_{*}^{2} \right) }{4r_{s}^{4}}-\frac{r_{*}^{2}}{r_{h}^{2}} \right] r_{h}^{2} \, r^{2}+ r_{*}^{2} \, r_{h}^{2}}}  \, ,
\nl
\Upsilon&=& \left[ \frac{1}{r_{h} l} \ln \frac{1-\frac{r_{h}^{2} \left( r_{s}^{2}-r_{*}^{2} \right) }{4r_{s}^{4}}+\frac{r_{*}^{2}}{r_{h}^{2}} - 2 \frac{r_{*}}{r_{h}}}{1-\frac{r_{h}^{2} \left( r_{s}^{2}-r_{*}^{2} \right) }{4r_{s}^{4}}+\frac{r_{*}^{2}}{r_{h}^{2}} + 2 \frac{r_{*}}{r_{h}}} + 2  \right] \, .
\eea
At late times we can use the following approximation, which can be derived from
eq. (\ref{espansion-late-time}):
\beq
r_{m} \approx \sqrt{\hat{r}_* r_h} + \sqrt{\epsilon_*} \sqrt{\frac{r_h^3 (r_h^2-2 r_s^2) }{8 r_s^3}} 
+\mathcal{O}(\epsilon_*)\, ,
\eeq
where we have used the property $r_s>r_h/2$, which is always valid.

The calculation of the various term proceeds as follows:
\begin{itemize}
\item
Let us focus on  $\mathcal{I}_4$.
Due to the Heaviside $\theta$, this term is non vanishing
just in the intermediate time window eq. (\ref{tempaccio}), i.e.
\beq 
 t< \frac{ l}{2} - \frac{0.53}{r_h} \, .
\eeq
Using the expansion in eq. (\ref{espansion-late-time}), we can approximate
\beq
\Upsilon = 2+ \frac{1}{r_h l} \ln \frac{\epsilon_* (r_h^2-2 r_s^2) r_h^2}{8 \hat{r}_* r_s^3} +\mathcal{O}(\epsilon_*) 
\approx
1+\frac{1}{r_h l} \le \frac{4(1-\rho_s)}{2 \rho_s-1} +\ln \frac{(1-\rho_s)(1-2 \rho_s^2)}{\rho_s (1-2 \rho_s +2 \rho_s^2)}
 \ri  \, ,
\eeq
where we have used the late time approximation in eq. (\ref{eppi-appi}).
We can also use the approximation:
\beq
\psi(r)= r \, \frac{\sqrt{r^2-\hat{r}_*^2}}{\sqrt{(r^2-\hat{r}_* r_h)^2 + \mathcal{O}(\epsilon_*)}} \, .
\eeq
This gives:
\beq
\mathcal{I}_4 \approx \theta \left( \frac{r_{h}}{\sqrt{2}} - r_{s} \right) l \,  \Upsilon
\int_{r_{m}}^{r_{s}} dr \, 
 r \, \frac{\sqrt{r^2-\hat{r}_*^2}}{r^2-\hat{r}_* r_h} \, .
\eeq
This integral has a cutoff at $r_m\approx \sqrt{r_h \hat{r}_*} +\mathcal{O}(\sqrt{\epsilon_*})$,
and so it is a good approximation to drop the order $\epsilon_*$ term in the denominator.
This can now be evaluated analytically, using:
\beq
\eta(r)= \int  dr \, r
 \frac{  \sqrt{ r^2-\hat{r}_{*}^2 }}{ (r^2-\hat{r}^* r_h)   }  =\sqrt{r^2-\hat{r}_*^2}
 +\frac12 \sqrt{(r_h-\hat{r}_*) \hat{r}_*} 
 \ln \left| \frac{\sqrt{(r_h-\hat{r}_*) \hat{r}_*}-\sqrt{r^2-\hat{r}_*^2}}
 {\sqrt{(r_h-\hat{r}_*) \hat{r}_*}+\sqrt{r^2-\hat{r}_*^2}} \right| \, .
 \label{primitiva-eta}
\eeq
We finally get:
\beq
\mathcal{I}_4 \approx 
\theta \left( \frac{r_{h}}{\sqrt{2}} - r_{s} \right) 
l \,  \Upsilon \le \eta(r_s)-\eta(r_m) \ri \, .
\eeq


\item
Let us consider $\mathcal{I}_2$, which at large time can be approximated as:
\beq
\mathcal{I}_2
\approx l \int_{r_{s}}^{\Lambda} dr \, r
  \sqrt{ \frac{ r^2-\hat{r}_{*}^2 }{ (r^2-\hat{r}_* r_h)^2 + \epsilon_* A(r)
 } } \, , \, \qquad A(r)= \frac{  \hat{r}_* \left(r^2 \left(4 r_s^4 r_h^2+r_h^6\right)-4 r_s^4 r_h^4\right)}{2 r_h  r_s^4} \, .
\eeq
It is  useful to use the following properties:
\beq
\begin{cases}
{\rm for}\,\,\,  r_s=\frac{r_h}{\sqrt{2}} \,\,\, {\rm and}\,\,\, r_s=r_h \, , \qquad & r_s=\sqrt{\hat{r}_* r_h}  \\
{\rm for}\,\,\,  r_s<\frac{r_h}{\sqrt{2}} \, , \qquad & r_s> \sqrt{\hat{r}_* r_h} \\
{\rm for}\,\,\,  r_h > r_s>\frac{r_h}{\sqrt{2}} \, , \qquad &  r_s< \sqrt{\hat{r}_* r_h} 
\end{cases}
\eeq
For this reason, we should separate two cases:
\begin{itemize}
\item For $r_s<\frac{r_h}{\sqrt{2}}$ we have that  $r_s> \sqrt{\hat{r}_* r_h} $
and so the $\epsilon_*$ term at the denominator is negligible:
\beq
\mathcal{I}_2 \approx \theta \left( \frac{r_{h}}{\sqrt{2}} - r_{s} \right)  l \, (\Lambda-\eta(r_s)) \, .
\eeq
\item For $r_s>\frac{r_h}{\sqrt{2}}$ it is convenient to split
\bea
\mathcal{I}_2 &=&\mathcal{I}_2^a +\mathcal{I}_2^b \, , \qquad
\mathcal{I}_2^a=
l \int_{r_s}^{\sqrt{\hat{r}_* r_h}}  dr \, r
 \frac{  \sqrt{ r^2-\hat{r}_{*}^2 }}{ \sqrt{(r^2-\hat{r}_* r_h)^2 + \epsilon_* A}   }  \, , 
 \nl
\mathcal{I}_2^b &=& l  \int_{\sqrt{\hat{r}_* r_h}}^{\Lambda}  dr \, r
 \frac{  \sqrt{ r^2-\hat{r}_{*}^2 }}{ \sqrt{(r^2-\hat{r}_* r_h)^2 +\epsilon_* A(r)}   }
\eea
We will not need to evaluate $\mathcal{I}_2^a$, because we will show that
it is cancelled by a term in $\mathcal{I}_3$.
We can approximate $\mathcal{I}_2^b$ by noting that the term proportional
to $\epsilon_*$ at the denominator acts  as an effective cutoff of the integral:
\beq
\mathcal{I}_2^b  \approx  l  \int_{\tilde{r}}^{\Lambda}  dr \, r
 \frac{  \sqrt{ r^2-\hat{r}_*^2 }}{ (r^2-\hat{r}_* r_h)}   \, , \qquad
 \tilde{r}=\sqrt{r_* r_h + \sqrt{\epsilon_* A(\sqrt{\hat{r}_* r_h})}} \, .
\eeq
Using eqs. (\ref{primitiva-eta}) and (\ref{eppi-appi-2}) we find the leading $l$
behaviour:
\beq
\mathcal{I}_2^b \approx
l \Lambda + \frac{\sqrt{r_s r_h \left(-2 r_s^2+3 r_s r_h - r_h^2 \right)}}{4 \left(2 r_s^2-2 r_s r_h +r_h^2 \right)} r_h^2 l^2 + \mathcal{O}(l) \, .
\eeq

\end{itemize}

\item
We now approximate $\mathcal{I}_3$.
In the limit $r_* \rightarrow \hat{r}_*$, we find that:
\beq
\kappa(r)=\frac{1}{r_h }  \ln \left(\frac{2 \rho^2 \rho_s^2-2 \rho^2 \rho_s
   +\rho^2-2 \rho_s^2+\rho_s +
\left| \rho^2 \left(2 \rho_s^2-2 \rho_s+1\right)-2 \rho_s^2+\rho_s\right|}
   {2 \rho^2 \rho_s^2-2 \rho^2 \rho_s
   +\rho^2+2 \rho_s^2-\rho_s
     + \left|(\rho^2 \left(2 \rho_s^2-2   \rho_s+1\right)-2 \rho_s^2+\rho_s\right| }\right)\, ,
\eeq
where we introduced $\rho=r/r_h$.
It is useful to consider separately the following two cases:
\begin{itemize}
\item
If $r_s<r_h/\sqrt{2}$, then
\beq
\rho^2 \left(2 \rho_s^2-2 \rho_s+1\right)-2 \rho_s^2+\rho_s >0
\label{roro}
\eeq
for every $r_s<r<\Lambda$ and the factor in the integrand
 is finite and  suppressed in the large $l$ limit.
\item
If $r_s>r_h/\sqrt{2}$, then eq. (\ref{roro})
is valid just for $r>\sqrt{r_* r_h}$, and again gives a negligible contribution.
For $r<\sqrt{r_* r_h}$ we have to change a sign
and we get that the factor in the integrand is:
\beq
\kappa(r)=\frac{1}{r_h } \ln \left(-\frac{\epsilon_*  \left(\rho^2 \left(4 \rho_s^5-2 \rho_s^3+2 \rho_s-1\right)
+2 \left(1-2 \rho_s\right)
   \rho_s^4\right)}{4 \rho_s^4 \left(2 \rho_s-1\right) \left(\rho^2 \left(2 \rho_s^2-2 \rho_s+1\right)-2
   \rho_s^2+\rho_s\right)}\right) \, .
\eeq
Inserting $\epsilon_*$ from the solution in eq. (\ref{eppi-appi}), we find that
the log  factor in the integrand simplifies to $-l+\mathcal{O}(l^0)$,
which cancels $\mathcal{I}_2^a $.
\end{itemize}
\end{itemize}

Adding up all the contributions,
we find eq. (\ref{tempi-medi}) for $r_s<r_h/\sqrt{2}$
and eq. (\ref{tempi-lunghi}) for $r_s>r_h/\sqrt{2}$.

\end{document}